\documentclass{aa}

\usepackage{graphicx}
\usepackage{multirow}
\usepackage{xcolor}
\usepackage{bm}
\usepackage{pdfpages}
\usepackage{txfonts}
\usepackage{soul} % to use st
\newcommand{\epscha}{$\epsilon$~Cha}
\newcommand{\cha}{Chamaeleon}
\newcommand{\Gaia}{\emph{Gaia}}

\def\new#1{{#1}}
\def\newst{\relax}

\begin{document}

   \title{New low-mass members of Chamaeleon I and $\epsilon$ Cha}

   \author{K. Kubiak\inst{1} , K. Mu\v{z}i\'c\inst{1}, I. Sousa\inst{1}, V. Almendros-Abad \inst{1},
   R. K\"ohler,\and A. Scholz\inst{2}
          %\fnmsep
          %\thanks{Just to show the usage
          %of the elements in the author field}
          }

   \institute{CENTRA, Faculdade de Ci\^{e}ncias, Universidade de Lisboa, Ed. C8, Campo Grande, P-1749-016 Lisboa, Portugal\\
              \email{karolina.kubiak@sim.ul.pt}
         \and
        {SUPA, School of Physics \& Astronomy, University of St Andrews, North Haugh, St Andrews KY169SS, UK}\\
             }

   \date{Received; accepted}

% \abstract{}{}{}{}{}
% 5 {} token are mandatory

  \abstract
   {}
   {The goal of this paper is to increase the membership list of the \cha~ star forming region and the \epscha~ moving group, in particular for low-mass stars and substellar objects. We extended the search region significantly beyond the dark clouds.}
   {Our sample has been selected based on proper motions and colours obtained from \Gaia\ and 2MASS. We present and discuss the optical spectroscopic follow-up of 18  low-mass stellar objects in Cha~I and \epscha. We characterize the properties of objects by deriving their physical parameters, both from spectroscopy and photometry.}
  {We add three more low-mass members to the list of Cha I, and increase the census of known \epscha\ members by more than 40\%, confirming spectroscopically 13 new members and relying on X-ray emission as youth indicator for 2 more.
  In most cases the best-fitting spectral template is from objects in the TW Hya association, indicating that \epscha\ has a similar age.
  The first estimate of the slope of the initial mass function in \epscha\ down to the sub-stellar regime is consistent with that of other young clusters. We estimate our IMF to be complete down to $\approx 0.03$\, M$_{\odot}$.  The IMF can be represented by two power laws: for M $<$ 0.5 M$_{\odot}\,\, \alpha = 0.42 \pm 0.11$ and for M $>$ 0.5 M$_{\odot} \,\,\alpha = 1.44 \pm 0.12$.
  }
   {We find similarities between \epscha\ and the southernmost part of Lower Centaurus Crux (LCC A0), both lying at similar distances and sharing the same proper motions.
   This suggests that \epscha~ and LCC A0 may have been born during the same star formation event}


  %  \titlerunning{Cha I and \epscha}
    \authorrunning{K. Kubiak et al.}
   \maketitle
%
%-------------------------------------------------------------------

\section{Introduction}
The \cha~ molecular cloud complex is one of the nearest sites of star formation and dominates the dust extinction maps of the region \citep{luhman08}. The complex and its surroundings are abundant with pre-main-sequence stars of various ages.  The large population of young stellar objects (YSOs) is associated with the Cha I \citep{luhman04a,luhman07}, Cha II \citep{spezzi08} and Cha III clouds (Fig.\,\ref{fig:area}). Because the \cha~cloud complex is nearby (Cha~I $d\approx190$\,pc, Cha~II $d\approx210$\,pc, \citealt{Dzib18}, \citealt{Zucker20}) and well-isolated from other young stellar populations, it has been a popular target for studies of low-mass star formation.  Cha~I harbours a rich population of $\sim$250 known YSOs with ages of $\sim$2\,Myr, grouped in two sub-clusters \citep{luhman07, luhmanmuench08, muzic11, sacco17}. The \cha~II  dark cloud, at a distance of 210 pc \citep{Zucker20}, has modest star formation activity with a population of 51 confirmed members aged 2 -- 4\,Myr \citep{spezzi08}. No young stellar objects are known in the dark cloud Cha III.

Between the clouds (in projection), we find a more nearby ($d\sim100$pc) and somewhat older (5 -- 10 Myr) group of YSOs, known as the $\epsilon$ \cha tis Association \citep[\epscha,][]{feigelson03,luhman08,Dzib18}. Once thought to be the result of ejection from the \cha~complex \citep{sterzik_durisen95}, many of these isolated, \textit{older} pre-main-sequence stars are now believed to form a distinct nearby young co-moving group of \epscha~  \citep{murphy13}.

One of the primary objectives of the survey presented in this paper is the extension of previous censuses of low mass stars and substellar objects in Cha I, Cha II and \epscha.  To achieve this, we use data from \Gaia\ DR2 \citep{GaiaDR2}.  \Gaia\ data has already been used in combination with near- and mid-infrared surveys to search for new low-mass stars and brown dwarfs in the solar neighbourhood (e.g. \citealt{esplin_luhman19}, \citealt{Canovas19}).

We confirmed the young age of candidate low-mass objects in Cha I, Cha II and \epscha\ by performing optical spectroscopy.
Analyzing low-resolution optical spectra is not only a powerful tool to confirm youth of low mass stellar objects, but also useful for investigating their physical properties by comparing with spectra of standard stars with well-defined properties and templates.  For nearby star-forming regions and associations, such observations can be performed with modest size telescopes. We performed spectroscopic follow-up using the FLOYDS instrument installed at the 2-m Faulkes South telescope, which is part of the Las Cumbres Observatory Global Network (LCOGT).

In this paper, we first describe the photometric datasets used for the analysis (Sect.~\ref{sec:data}) and introduce the known members (Sect.~\ref{sec:members}), followed by the selection of candidates from density maps, proper motions, and colour-magnitude diagrams (Sect.~\ref{sec:regions}). In Sect.~\ref{sec:follow-up}, we present the analysis of the optical spectra.  We then characterize the stellar parameters, spatial distribution, and spectral energy distributions of the confirmed members and summarize their properties. In the Appendix, we present the spectral template fits to FLOYDS spectra obtained in this study.

%%%%%%%%%%%%%%%%%%%%%%%%%%%%%%%%%%%%%%%%%%%%%%%%%%%%%%%%%%%%%%%%%%%%%%%%%%%
\section{Datasets}
\label{sec:data}
\begin{figure*}
   \centering
   \includegraphics[width=0.9\textwidth]{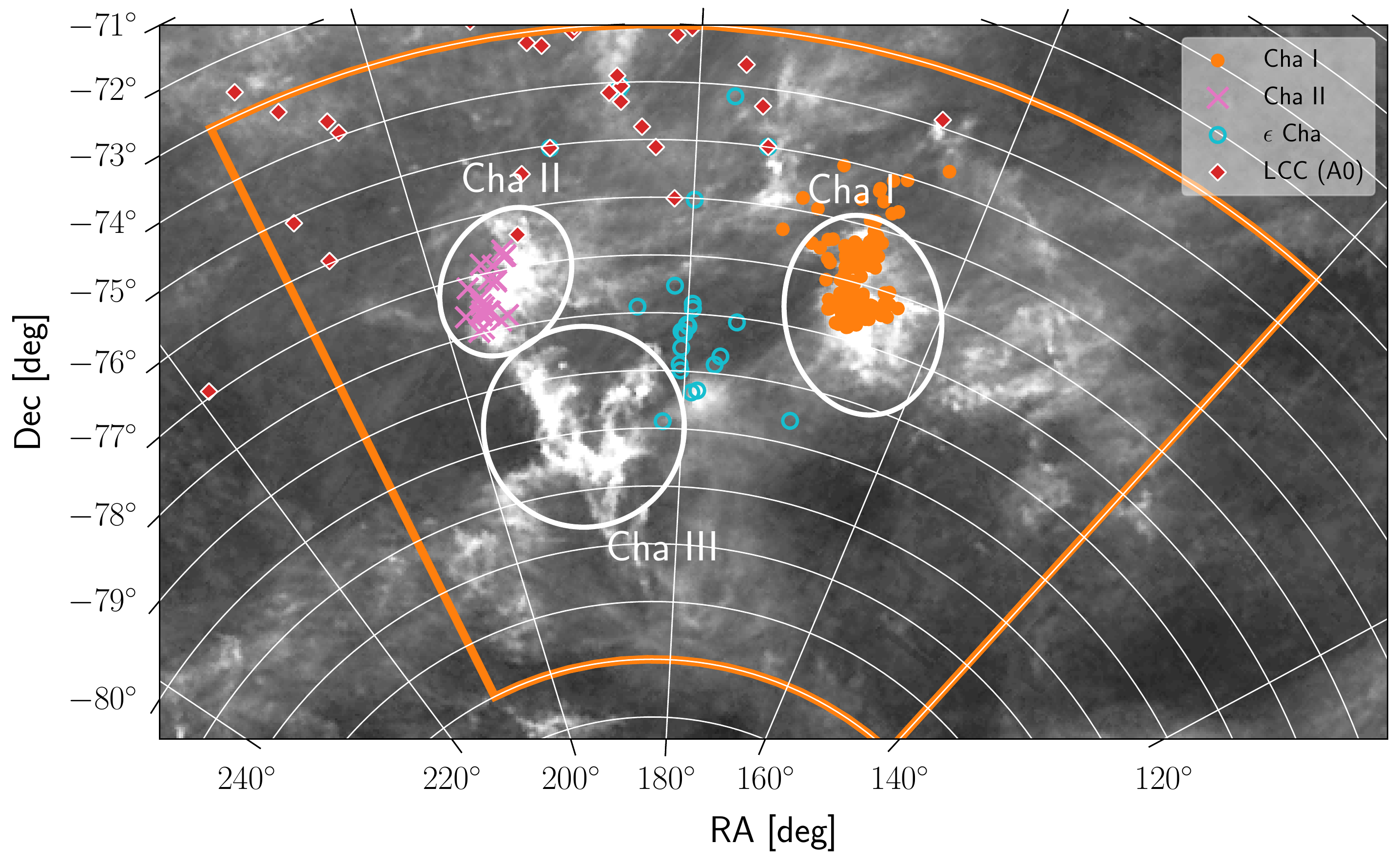}
      \caption{$Planck$ 857 GHz dust map of the \cha~ region. The positions of the main dark clouds are marked with white ellipses. The area studied in this work is encompasses by the orange lines. The known members of Cha~I and Cha~II star-forming regions are shown as filled orange circles and pink crosses, respectively \citep{luhman07, luhmanmuench08, muzic11,sacco17,spezzi08}. The open cyan circles mark the \epscha~ members \citep{murphy13}, and red diamonds are kinematic candidates associated with the A0 group of LCC from \citet{goldman18}.
              }
         \label{fig:area}
   \end{figure*}

The region analysed in this work is located in the coordinate range $140^{\circ} < \alpha < 210^{\circ}$, and $-84^{\circ} < \delta < -73^{\circ}$.
This large area, as shown in Fig.\,\ref{fig:area}, encompasses the \cha~clouds, and the \epscha~moving group. The catalogues described in this section were cross-matched using TOPCAT \citep{topcat}, with a matching radius of 1$''$.

%-----------------------------------------------------------------

\subsection{Gaia DR2}

The Gaia \citep{GaiaMission} Data Release 2 (DR2; \citealt{GaiaDR2}) catalog of the region was queried through VizieR \citep{vizier}. The catalogue was restricted to parallaxes between 3.3 and 20\,mas (equivalent to distances 50-300 pc), in order to reduce its size, but at the same time include the distances of interest, with a generous margin. Furthermore, we excluded the objects with only one good observation, as reported by the catalogue keyword $astrometric\_n\_good\_obs\_al$.
%, and those with RUWE$<3$\footnote{\url{https://dms.cosmos.esa.int/COSMOS/doc_fetch.php?id=3757412}}, to assure good astrometric quality. We note that the threshold value for RUWE is slightly higher than the often adopted value of 1.4. We inspected the RUWE parameter for the known members, and find that a significant portion of those has elevated RUWE values, possibly due to the presence of circumstellar material. We find that $\sim$90\% of the known members have RUWE < 3, and therefore decide to adopt this value.

\subsection{2MASS}
The near-infrared photometry was obtained from the Two Micron All Sky Survey (2MASS; \citealt{2mass}), queried through the Infrared Science Archive (IRSA) interface\footnote{\url{https://irsa.ipac.caltech.edu/}}. We kept only the photometric points with the quality flag ($ph\_qual$) flag values of A, B, or C for any of the three 2MASS bands ($J$, $H$, and $K_S$).

\subsection{AllWISE}

The mid-infared photometry (bands  $W1$, $W2$, $W3$, and $W4$, centred at
$3.6\mu$m, $4.6\mu$m, 11.6$\mu$m, and $22.1\mu$m, was obtained from the AllWISE project, making use of data from the Wide-field
Infrared Survey Explorer cryogenic survey (WISE; \citealt{WISE}). We have filtered out the photometry for the bands with signal-to-noise ratio below 5 (keywords $w1snr$, $w2snr$, $w3snr$ and $w4snr$). Furthermore, the following flag values have been accepted for $W1$ and $W2$ photometry:
$cc\_flags$ = 0,
$ext\_flg$ = 0 - 3,
$ph\_qual$ = A or B, and
$w1flg$ = $w2flg$ = 0.

\subsection{Known members}
\label{sec:members}
The principal young populations known to be present  towards the surveyed region are Cha I, Cha II and \epscha. In the north direction, at distances similar to \epscha, we find Lower-Centaurus-Crux (LCC), part of the large scale Sco-Cen OB association \citep{dezeeuw99}. In the original work by \citet{dezeeuw99} the LCC extends northward of the galactic longitude b$>-10^{\circ}$, corresponding to $\delta \gtrsim -70^{\circ}$, i.e. outside the region studied here. The same is valid for the previous spectroscopic studies of the LCC candidate objects by \citet{mamajek02, song12}. However, a recent study based on Gaia DR2 astrometry \citep{goldman18}, searched for young sources over a slightly wider region, in which they subdivide the candidate members in four subgroups, characterised by the increasing isochronal age from 7 to 10 Myr. The youngest group, labelled A0 by \citet{goldman18}, partially overlaps with our search area, and several of its candidate members have been earlier recognised as members of \epscha~ \citep{murphy13}. We mark the candidate members from \citet{goldman18} as "LCC" in Fig.\,\ref{fig:area}, noting that some of these objects may as well be members of \epscha. As we will discuss in more details later, they share the same proper motion and parallax space, as well as the location in the barycentric Galactic Cartesian coordinate frame. For this reason also, we do not use Goldman et al. candidates for the new member selection, we rather rely on \epscha~members, most of which have additionally been confirmed by spectroscopy.

For our analysis, we take into account the following lists of members:

\begin{enumerate}
    \item Cha I: A list of 256 spectroscopically confirmed members has been compiled using the works from \citet{luhman07, luhmanmuench08, muzic11, sacco17}.

    \item Cha II: A list of 51 spectroscopically confirmed members published in \citet{spezzi08}.

    \item \epscha: A comprehensive list of 35 members has been compiled by \citet{murphy13}, using both kinematic and spectroscopic information.
\end{enumerate}
The previously confirmed members, along with the LCC (A0) candidates are shown in Fig.\,\ref{fig:area}.

%Gaia data reveal that some of the sources previously identified as members show parallaxes that are discrepant with the membership in the above mentioned groups. The histograms with the putative member parallaxes are shown in Figure~\ref{fig:?}. The objects outside of the region marked by the {\ color {red} dashed lines} have been excluded from the member lists.
%Note: this is not necessary as these sources will automatically be discarded in the candidate selection process, which only takes the sources that match our own catalogue, restricted by the parallax.
%%%%%%%%%%%%%%%%%%%%%%%%%%%%%%%%%%%%%%%%%%%%%%%%%%%%%%%%%%%%%%%%%%%%%

\section{Candidate selection}
\label{sec:regions}
\subsection{On-sky density maps}

   \begin{figure}
   \centering
   \includegraphics[width=0.47\textwidth]{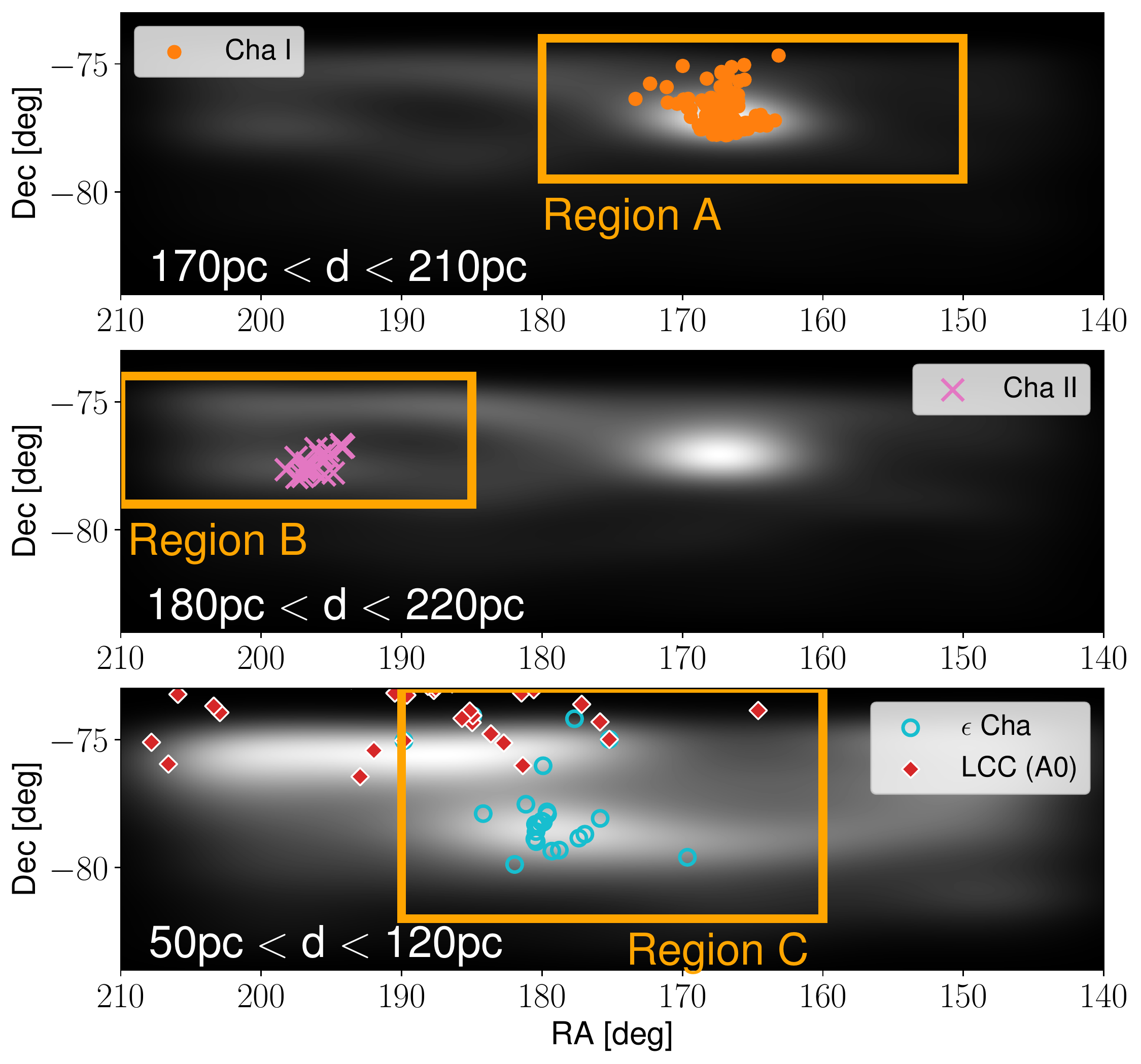}
      \caption{$J$-band KDE maps for three different distance cuts. The three studied regions are marked (see Section~\ref{sec:regions}), along with the known members of various clusters and associations found in the area. The symbols are the same as in Fig.\,\ref{fig:area}.}
         \label{fig:regions}
   \end{figure}

To examine the large region shown in Fig~\ref{fig:area}, we first plot stellar surface density maps limiting each map with a range of distances obtained from Gaia.
The kernel density estimations (KDE) of the positions in the 2MASS
$J-$band, were plotted for the distance range of 50 pc to 300 pc, with
steps of 10 pc. The KDEs were obtained using a Gaussian kernel with a
bandwidths of 25$'$, 25$'$ and 31$'$, for 3-dimensional (RA, Dec,
parallax) for regions A to C, respectively\footnote{The lower density
  at the edges of the plots is an artefact. KDE sums Gaussian
  contributions from each star and regions close to the edges miss the
  contribution from the sources located beyond them.}. This allows us
to visually identify stellar surface overdensities associated with the
groups of interest in which we aim to select new candidates. After the
visual inspection,{\newst {we decide to focus on the following cubes
  where we investigated the most prominent stellar overdensities
  associated with regions}} {\new we decided  to focus on the most prominent
overdensities associated with the regions}:

\begin{enumerate}
    \item Region A (associated with the Cha I cloud):\\
    $150^{\circ} < \alpha < 180^{\circ}$\\
    $-79.5^{\circ} < \delta < -74.5^{\circ}$\\
    4.75\,mas < $\varpi$ < 5.9 \,mas ($\approx$ 170 pc < d < 210 pc)
    \vspace{0.1 cm}
    \item Region B (associated with the Cha II cloud):\\
    $185^{\circ} < \alpha < 210^{\circ}$\\
    $-79^{\circ} < \delta < -74^{\circ}$\\
    4.55\,mas < $\varpi$ < 5.65 \,mas ($\approx$ 180 pc < d < 220 pc)
    \vspace{0.1 cm}
    \item Region C (in front of \cha\ clouds,
    associated with \epscha~and partially with LCC):\\
   % $155^{\circ} < \alpha < 190^{\circ}$\\
   % $-82^{\circ} < \delta < -76.5^{\circ}$\\
    $160^{\circ} < \alpha < 190^{\circ}$\\
    $-82^{\circ} < \delta < -73^{\circ}$\\
    8.3\,mas < $\varpi$ < 20 \,mas ($\approx$ 50 pc < d < 120 pc).
\end{enumerate}

The density maps showing these three regions are shown in Fig.\,\ref{fig:regions}, along with the known members described in Section~\ref{sec:members}.
%Note that the manipulation of the catalogues in our selection is based on the parallax, and not the distance, in order to avoid the issues caused by simple inversion of the parallax \citep[e.g.][]{bailerjones18}.

{\new The 3D areas from which regions A and B were selected overlap significantly (as shown in Fig. 2, the coordinate ranges are identical, but both cubes are slightly shifted in parallax range)}, and we restricted our study to the vicinity of
Cha II in the second {\newst{cube}} {\new 3D area} (middle panel in Fig.\,\ref{fig:regions}),
even though it is not visually the most prominent. The lower panel in
Fig.\,\ref{fig:regions} shows the most complex and extended
enhancements in stellar densities. The two elongated overdensities are
associated with previously identified young moving group \epscha~and
recently postulated group A0 form LCC from \citet{goldman18}. The two
density enhancements seem to be separated by a strip of lower surface
density. We will discuss the relation between the two apparent
structures in Sec.~\ref{sec:regionC}. {\new The elongated shape of the
  overdensities is a projection effect due to the large size of the
  region and its far south location as shown in Fig.\,\ref{fig:area}. The
  purpose of presenting KDE maps here is only to guide our choice of
  the borders of regions (in $\alpha$ and $\delta$) from which we aim to select new candidates. }

\subsection{Proper motion and colour selection}
\label{sec:selection}

\begin{figure*}[htb]
    \centering
    \includegraphics[width=0.9\linewidth]{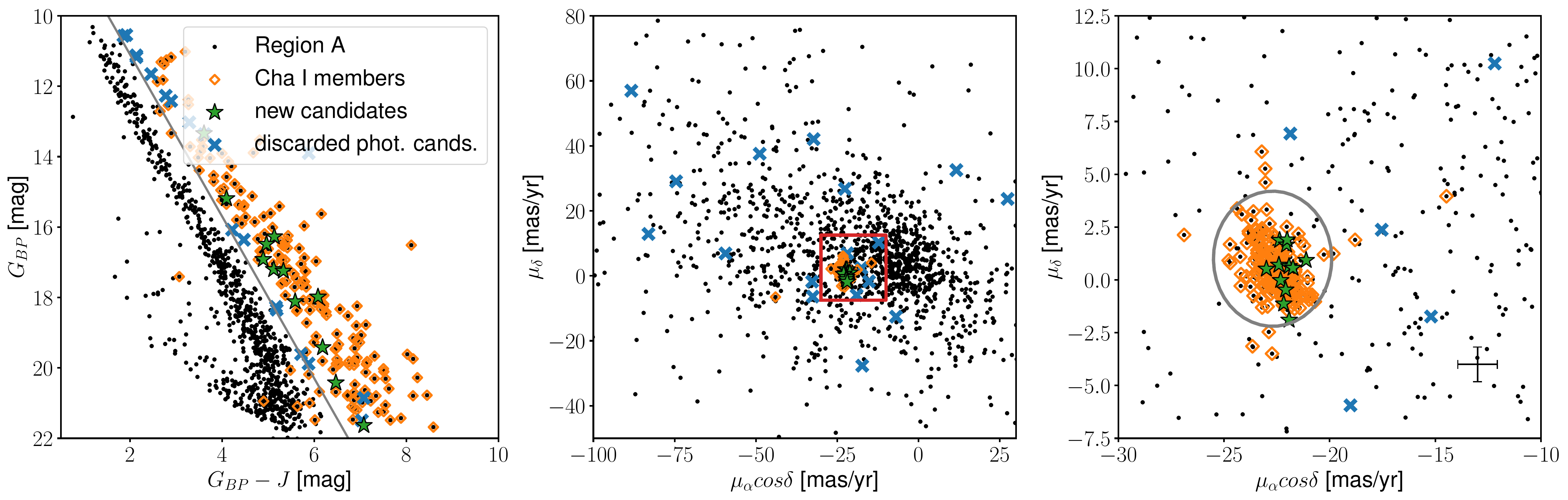}%

    \includegraphics[width=0.9\linewidth]{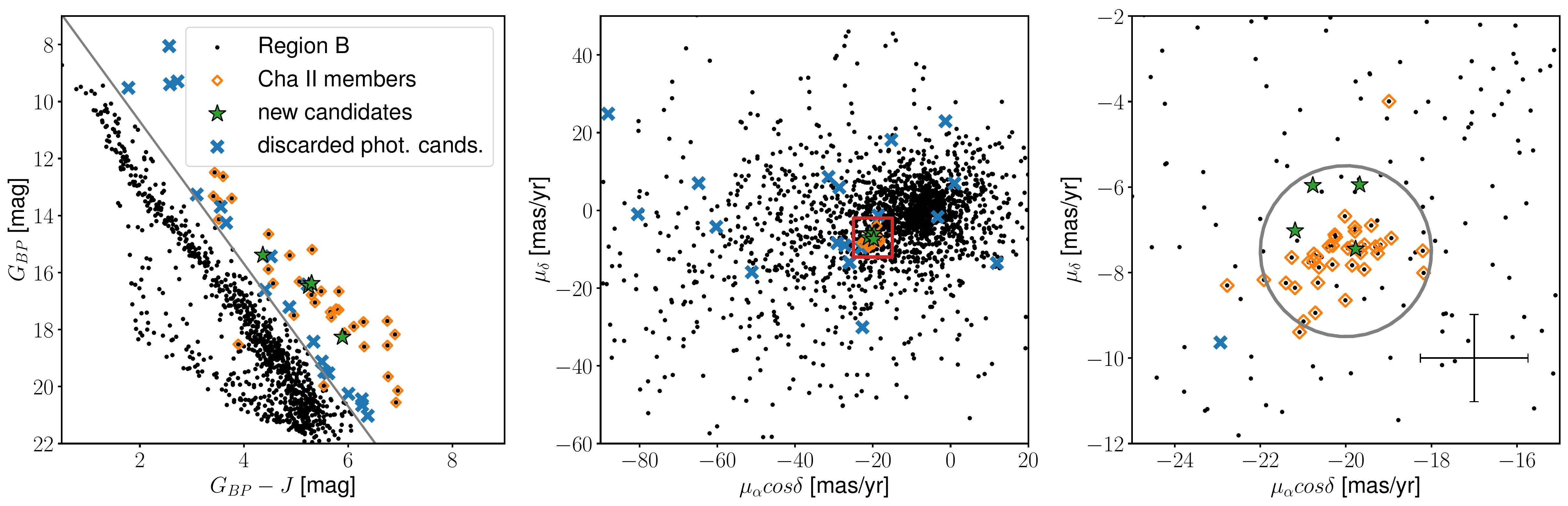}%

    \includegraphics[width=0.9\linewidth]{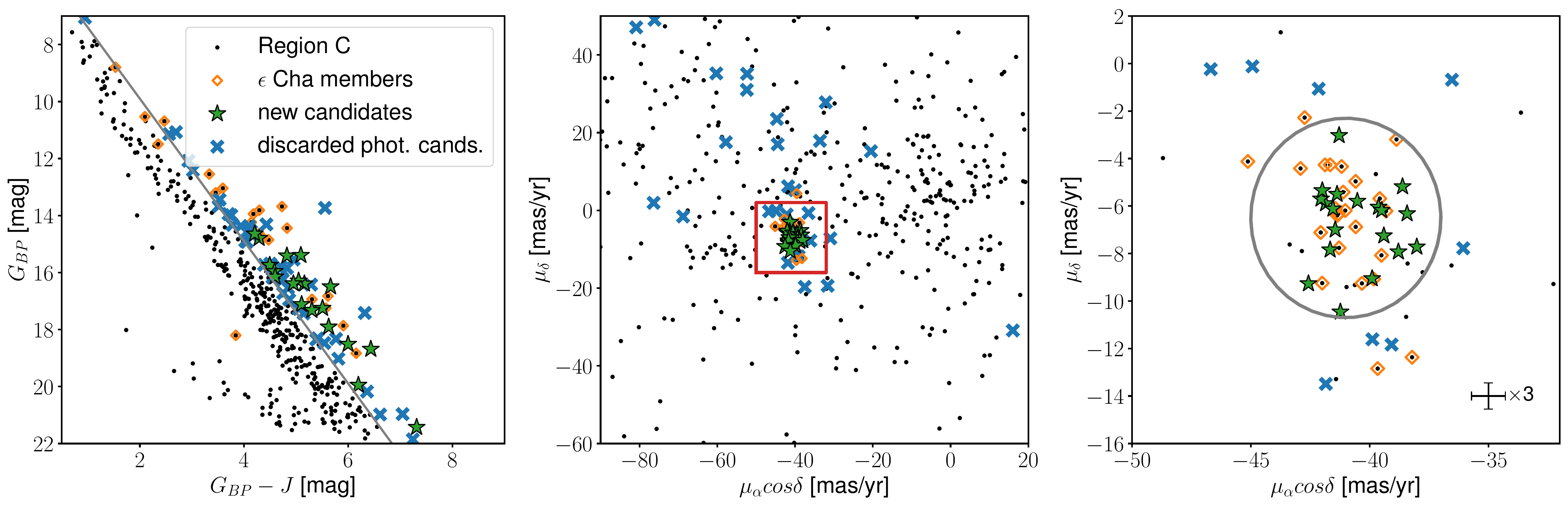}%
    \caption{Candidate selection based on colours and proper motions, with each row corresponding to one of the three regions of interest. The panels to the left show the $G_{BP} - J$, $G_{BP}$ CMDs. The middle panels contain the proper motions, with the red rectangle marking the zoomed-in region of the plot, which is shown in the right-hand side panels. New candidates (green stars) are accepted if their colours and proper motions overlap with those of previously confirmed members (orange diamonds), i.e. they are both located to the right of the grey line in the panels to the left, and inside the grey ellipse shown in the right-hand panels. {\new Precise equations are listed in the Appendix \ref{sec:math}}. Photometric members with discrepant proper motions are marked with blue crosses. The average proper motion uncertainties are shown in the lower-right part of the zoomed-in proper motion plots; for the region C they were multiplied by 3 for better visibility.}
    \label{fig:selection}
\end{figure*}

\begin{figure}
   \centering
   \includegraphics[width=0.48\textwidth]{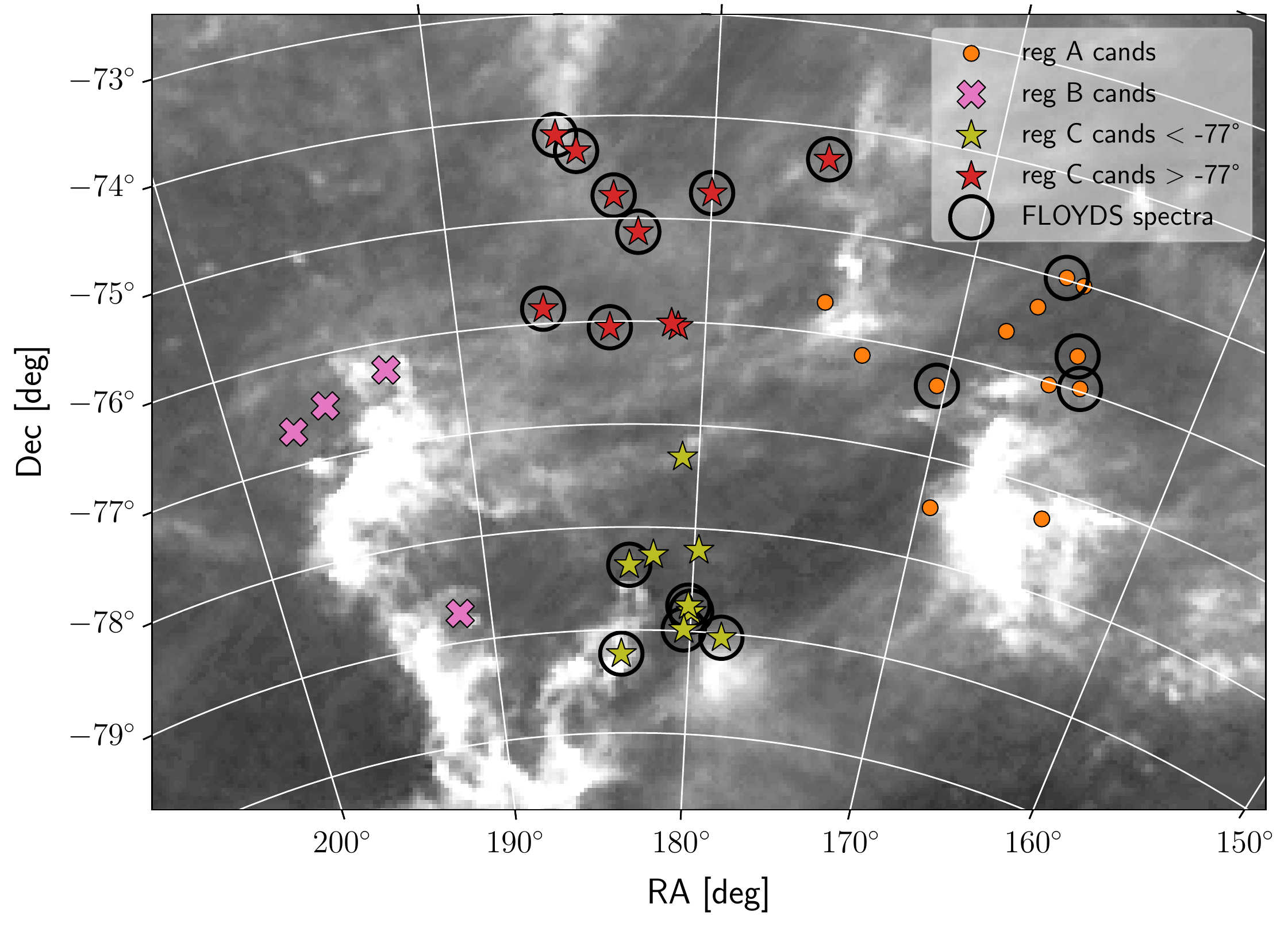}
      \caption{$Planck$ 857 GHz dust map of the \cha~ region, showing the candidates identified in this work. The black open circles mark those candidates for which we obtained spectra using FLOYDS spectrograph.
              }
         \label{fig:planck_cands}
   \end{figure}

The left column of Fig.\,\ref{fig:selection} exhibits the $G_{BP} -
J$, $G_{BP}$ colour-magnitude diagrams (CMDs) for the three regions
described in the previous section. The previously identified members
of Cha I, Cha II, and \epscha~(orange diamonds) are located to the
right of the main bulk of the sources, as expected from their pre-main
sequence status. They also cluster in the proper motion space (middle
and right panels in Fig.\,\ref{fig:selection}). To select new
candidates, we draw a line in the CMDs separating the field stars and
the red photometric candidates, as well as an ellipse in the proper
motion space, selected as to encompass most of the known members. {\new
 Our approach to the selection was a conservative one, aiming at a sample with less candidates, but with higher likelihood of membership. The exact functional forms are listed in the
Appendix \ref{sec:math}. } The sources that pass the two cuts, and are not on our previous member lists (Section~\ref{sec:members}), are considered new candidates for low-mass young members of Cha I, Cha II, \epscha, or LCC. The total number of the new candidates is 35: 12 in region A, 4 in region B, and 19 in region C. All the new candidates are listed in Table~\ref{tab:candidates} and shown in Fig.\,\ref{fig:planck_cands}, and their membership status will be discussed in more details in the following sections.

% \include{table_candidates}

%When pasting the new version of the table, add:
%table to table*
%move caption to the top
%\hline \hline after begin{tabular}
%\hline before end{tabular}
%after obj 12:14:04.54 & -79:13:52.0: & & & & & & & & & 2020-02-28 & 5x300s \\
%after obj 12:22:48.48 & -74:10:20.6: & & & & & & & & & 2020-05-03 & 4x600s\\
%after ID & RA... line add:	 & (hh:mm:ss) & (dd:mm:ss) & (mag) & (mag) & (mas) & (mas\,yr$^{-1}$) & (mas\,yr$^{-1}$) &  & & (s) \\
\begin{table*}
	\caption{The list of the new candidates. The final two columns refer to the FLOYDS follow-up.}
	\begin{center}
		\begin{tabular}{c c c c c c c c c c c}
		\hline \hline
			ID & RA & Dec & G$_{BP}$ & J & $\varpi$ & $\mu_{\alpha}$ & $\mu_{\delta}$ & reg. & date & $t_{\rm exp}$ \\
			  & (hh:mm:ss) & (dd:mm:ss) & (mag) & (mag) & (mas) & (mas\,yr$^{-1}$) & (mas\,yr$^{-1}$) &  & & (s) \\
			\hline
			1 & 10:58:17.95 & -77:17:19.9 & 18.12 & 12.53 & 5.41 & -22.37 &  1.88 & A &  &  \\
			2 & 10:58:54.74 & -75:58:19.6 & 16.27 & 11.14 & 5.07 & -22.40 &  0.70 & A &  2020-05-04 & 5x300 \\
			3 & 11:00:49.18 & -75:40:41.4 & 17.20 & 12.08 & 5.05 & -22.04 &  1.66 & A &  2020-05-05 & 3x600 \\
			4 & 11:03:02.73 & -75:00:30.1 & 21.63 & 14.56 & 5.20 & -22.01 &  1.90 & A &  &  \\
			5 & 11:03:51.51 & -76:01:32.4 & 20.42 & 13.95 & 4.98 & -21.92 &  0.64 & A &  &  \\
			6 & 11:05:51.57 & -74:58:39.6 & 15.19 & 11.10 & 5.12 & -21.73 &  0.56 & A &  2020-05-05 & 5x300 \\
			7 & 11:08:53.18 & -75:19:37.4 & 13.34 &  9.74 & 5.07 & -21.11 &  0.93 & A &  &  \\
			8 & 11:12:38.06 & -75:38:04.5 & 19.42 & 13.24 & 5.11 & -22.32 & -0.02 & A &  &  \\
			9 & 11:17:57.86 & -77:28:39.1 & 16.49 & 11.52 & 5.30 & -22.99 &  0.52 & A &  &  \\
			10 & 11:21:26.60 & -76:18:25.7 & 16.92 & 12.03 & 5.32 & -22.08 & -0.47 & A &  2020-05-05 & 3x600 \\
			11 & 11:34:20.34 & -76:09:10.2 & 17.25 & 11.92 & 5.34 & -22.17 & -1.13 & A &  &  \\
			12 & 11:41:24.64 & -75:41:41.0 & 17.98 & 11.90 & 4.90 & -21.92 & -1.90 & A &  &  \\
			13 & 12:46:05.97 & -78:43:03.7 & 16.47 & 11.23 & 5.08 & -20.77 & -5.96 & B &  &  \\
			14 & 12:52:18.62 & -76:16:00.6 & 16.38 & 11.08 & 5.12 & -21.19 & -7.02 & B &  &  \\
			15 & 13:03:16.08 & -76:29:38.2 & 18.27 & 12.38 & 4.97 & -19.68 & -5.94 & B &  &  \\
			16 & 13:09:26.07 & -76:40:13.9 & 15.39 & 11.02 & 4.98 & -19.77 & -7.45 & B &  &  \\
			17 & 11:43:29.51 & -74:18:37.9 & 15.74 & 11.25 & 10.06 & -41.28 & -3.03 & C &  2020-04-15 & 5x300 \\
			18 & 11:53:34.21 & -79:02:33.2 & 15.39 & 10.30 & 8.87 & -41.85 & -5.90 & C &  2020-04-18 & 3x300 \\
			19 & 11:59:07.79 & -78:12:32.1 & 17.12 & 12.01 & 9.42 & -38.61 & -5.18 & C &  &  \\
			20 & 12:00:00.90 & -78:48:29.0 & 15.41 & 10.59 & 9.68 & -41.37 & -5.50 & C &  2020-04-14 & 3x300 \\
			21 & 12:00:02.53 & -74:44:06.8 & 15.96 & 11.37 & 10.07 & -42.05 & -5.70 & C &  2020-04-19 & 4x300 \\
			22 & 12:00:37.72 & -78:45:08.4 & 18.51 & 12.52 & 9.77 & -41.54 & -6.12 & C &  2020-04-19 & 4x600 \\
			23 & 12:01:19.58 & -78:59:05.8 & 17.25 & 11.75 & 9.79 & -41.99 & -5.35 & C &  2020-04-13 & 7x300 \\
			24 & 12:02:54.45 & -77:18:38.2 & 14.80 & 10.51 & 9.61 & -39.63 & -6.04 & C &  &  \\
			25 & 12:04:24.99 & -76:02:42.3 & 21.43 & 14.11 & 9.56 & -40.52 & -5.79 & C &  &  \\
			26 & 12:05:29.12 & -76:00:52.3 & 19.95 & 13.76 & 9.40 & -39.49 & -6.18 & C &  &  \\
			27 & 12:07:45.78 & -78:16:06.6 & 16.15 & 11.55 & 9.45 & -38.42 & -6.32 & C &  &  \\
			28 & 12:10:54.28 & -75:07:54.9 & 16.39 & 11.23 & 9.72 & -39.40 & -7.25 & C &  2020-04-18 & 4x300 \\
			29 & 12:12:22.65 & -78:22:04.1 & 16.50 & 10.84 & 9.74 & -41.45 & -7.00 & C &  2020-04-19 & 3x300 \\
			30 & 12:14:04.54 & -79:13:52.0 & 16.29 & 11.25 & 9.87 & -41.66 & -7.84 & C &  2020-03-30 & 5x300 \\
			& & & & & & & & & 2020-02-28 & 5x300 \\
			31 & 12:14:34.00 & -74:46:41.0 & 16.39 & 11.44 & 9.52 & -38.79 & -7.93 & C &  2020-04-19 & 5x300 \\
			32 & 12:15:25.62 & -76:03:30.4 & 17.91 & 12.28 & 9.19 & -38.01 & -7.72 & C &  2020-05-11 & 3x600 \\
			33 & 12:19:53.45 & -74:20:09.5 & 14.64 & 10.44 & 9.82 & -39.89 & -9.05 & C &  2020-04-19 & 4x300 \\
			34 & 12:22:48.48 & -74:10:20.6 & 18.69 & 12.26 & 10.60 & -42.57 & -9.27 & C &  2020-05-11 & 3x600 \\
			& & & & & & & & & 2020-05-03 & 4x600\\
			35 & 12:25:58.09 & -75:51:11.8 & 17.32 & 12.02 & 10.05 & -41.23 & -10.46 & C &  2020-05-04 & 3x600 \\
			\hline
		\end{tabular}
	\end{center}
	\label{tab:candidates}
\end{table*}

   \subsection{Gaia eDR3}
  {\new  After the Gaia Early Data Release 3 (Gaia eDR3) on the 3
   of December 2020 we repeated the candidate selection based on
   eDR3, applying the identical selection conditions as described in previous sections.
   %The exact values of
   %the cut off are listed in Appendix\, \ref{sec:math}.
   The list of selected
   candidates remains unchanged for Cha II and \epscha, whereas one additional
   source passes our selection for the Cha I region.  This source
   has coordinates $\alpha$, $ \delta$ = 11:19:36.44, -75:08:32.8; parallax and proper motions
   $\varpi$ =5.84\,mas and ($\mu_{\alpha}$, $\mu_{\delta}$ ) =
   (-23.07, -1.10)\,mas\,yr$^{-1}$. An improvement in the proper motion measurements in eDR3
   causes the source to pass our selection criteria for the Cha I region. However, since
   the release
   of Gaia eDR3 occurred after the selection and analysis of the candidates, and follow
   up spectroscopy has been finished already we do not include this
   additional source in the analysis.}

   \subsection{Spectral energy distributions} %and HR-diagram}
   \label{sec:SED}

   \begin{figure*}
   \centering
   \includegraphics[width=\textwidth]{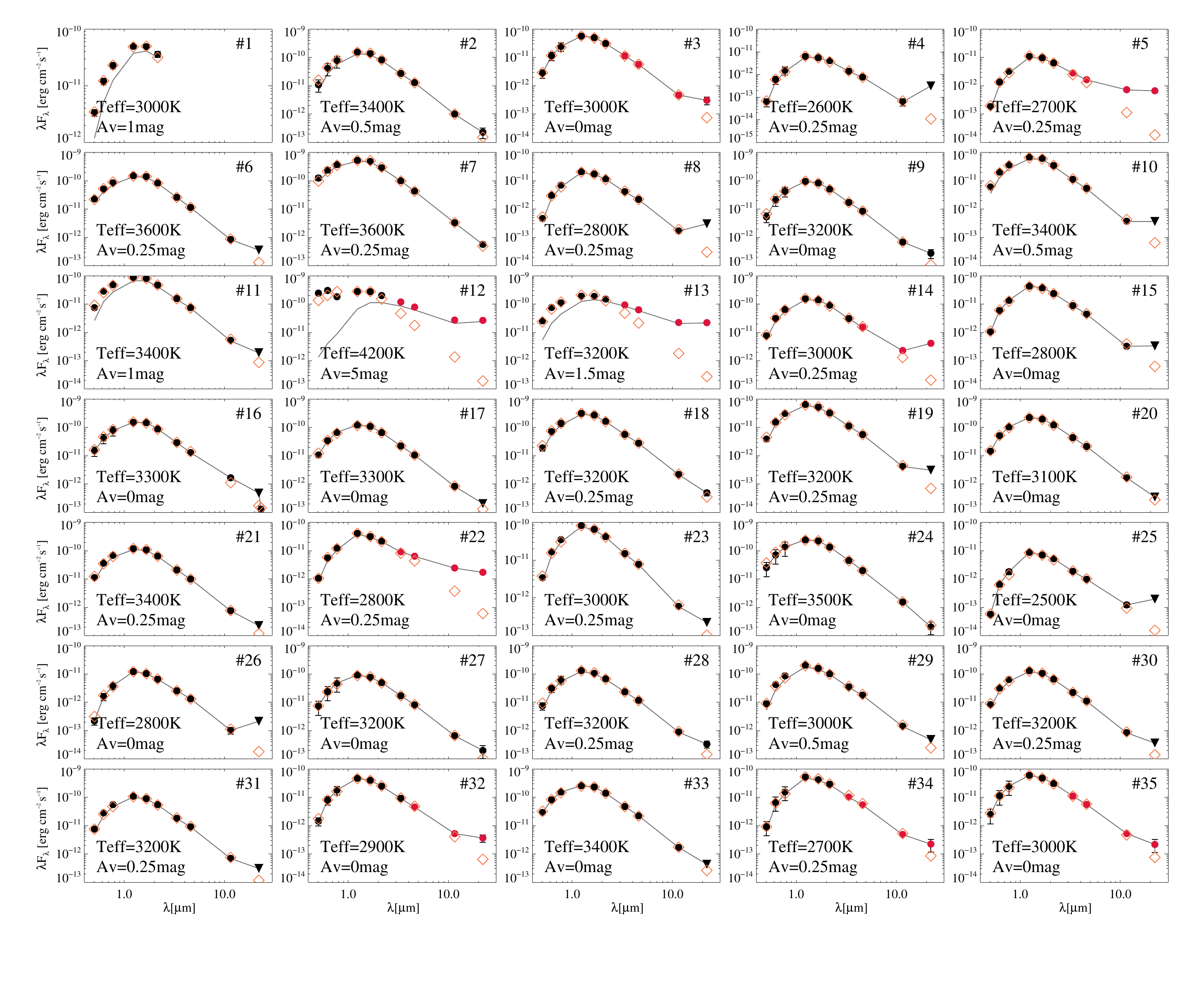}
      \caption{SEDs for all the identified candidates. The grey line shows the observed photometry. The filled circles connected with a grey line correspond to the observed photometry corrected from the best-fit value of the extinction, where the black circles are those used for the fitting, and the red ones were ignored due to excess emission at these wavelengths. The orange diamonds display the best-fitting BT-Settl model.
              }
         \label{fig:sed}
   \end{figure*}

The spectral energy distributions (SEDs) were constructed using the photometry from Gaia, 2MASS and WISE, and the SED fitting was performed with the help of the Virtual Observatory SED Analyzer (VOSA; \citealt{bayo08}). VOSA takes the fluxes and the distance for each object, and looks for the best fit effective temperature ($T_{\mathrm{eff}}$), extinction (A$_V$), and surface gravity (log$g$) by means of $\chi^2$ minimisation. For the objects showing excess in the WISE photometry, the SED fitting is performed over the optical and near-infrared portions of the SED, otherwise the full available range is included in the fit. The metallicity was fixed to the Solar value. We use the BT-Settl models \citep{allard11}, and probe $T_{\mathrm{eff}}$ over the full range offered in VOSA (400 - 7,000 K), with a step of 100 K, and A$_V$ between 0 and 5\,mag, with a step of 0.25\,mag (automatically determined by VOSA). The only exception is the candidate $\#12$, which is located in a small dusty cloud north-east of the main Cha I cloud, and requires larger values of A$_V$ to be considered (A$_V$=0-10\,mag). Furthermore, we allowed log$g$ to vary between 3.5 and 5.0, which encompasses the typical ranges for young cool very low-mass stars and brown dwarfs, and field stars. However, as previously noted by e.g. \citet{bayo17}, the SED fitting procedure is largely insensitive to log$g$, resulting in flat probability density distributions over the inspected range. Therefore the two main parameters constrained by this procedure are $T_{\mathrm{eff}}$ and A$_V$. The results of the SED fitting for all the candidates are given in the first three columns of Table~\ref{tab:spt_par}, and are shown in Fig.\,\ref{fig:sed}.

\section{Spectroscopic follow-up and analysis}
\label{sec:follow-up}
\subsection{Observations and data reduction}
Optical spectra of {\newst{the 18 brightest}} {\new 18} low-mass
candidates were taken using the FLOYDS spectrograph installed at the
2-m Faulkes South telescope, making part of the Las Cumbres
Observatory Global network (LCOGT). {\new
%We selected our targets starting from the brightest objects,
The priority in target selection was given to objects in region C,
aiming to observe the faintest possible candidates, while
maximising the efficiency and the total number of observed objects. In
the end of the observing cycle we supplemented our target list by 4 candidates from
region A in similar brightness range.}
The spectrograph provides a wide wavelength coverage of 3200 - 10000~\AA, with a spectral resolution R$\sim 400-700$. We set an upper wavelength cutoff at 9000\,\AA\ due to the strong telluric absorption present at longer wavelengths. Furthermore, a lower wavelength limit has been imposed on a portion of our spectra, due to a very low signal-to-noise ratio (SNR) caused by the faintness of cool objects in the blue part of the spectrum. We employed a slit of $1.6''$, and observed a flat field and an arc just before and after the science observations. For data reduction, we use the FLOYDS pipeline\footnote{\url{https://lco.global/documentation/data/floyds-pipeline/}}, which performs overscan subtraction, flat fielding, defringing, cosmic-ray rejection, order rectification, spectral extraction, and flux and wavelength calibration. The list of the observed objects, along with the dates and exposure times are given in Table~\ref{tab:candidates}. The data observed at two different dates for the object $\#34$ were combined to increase SNR, while we treat the two epochs of object $\#30$ separately in order to analyse the strongly variable H$\alpha$ emission. The candidates with spectra are marked with black circles in Fig.\,\ref{fig:planck_cands}.

%--------------------------------------------------------------------
%Table with the observation summary
%\include{table_observations}

%-------------------------------------------------------------------
%-----------------------------------------------------------------
   \begin{figure*}
   \centering
   \includegraphics[width=0.9\textwidth]{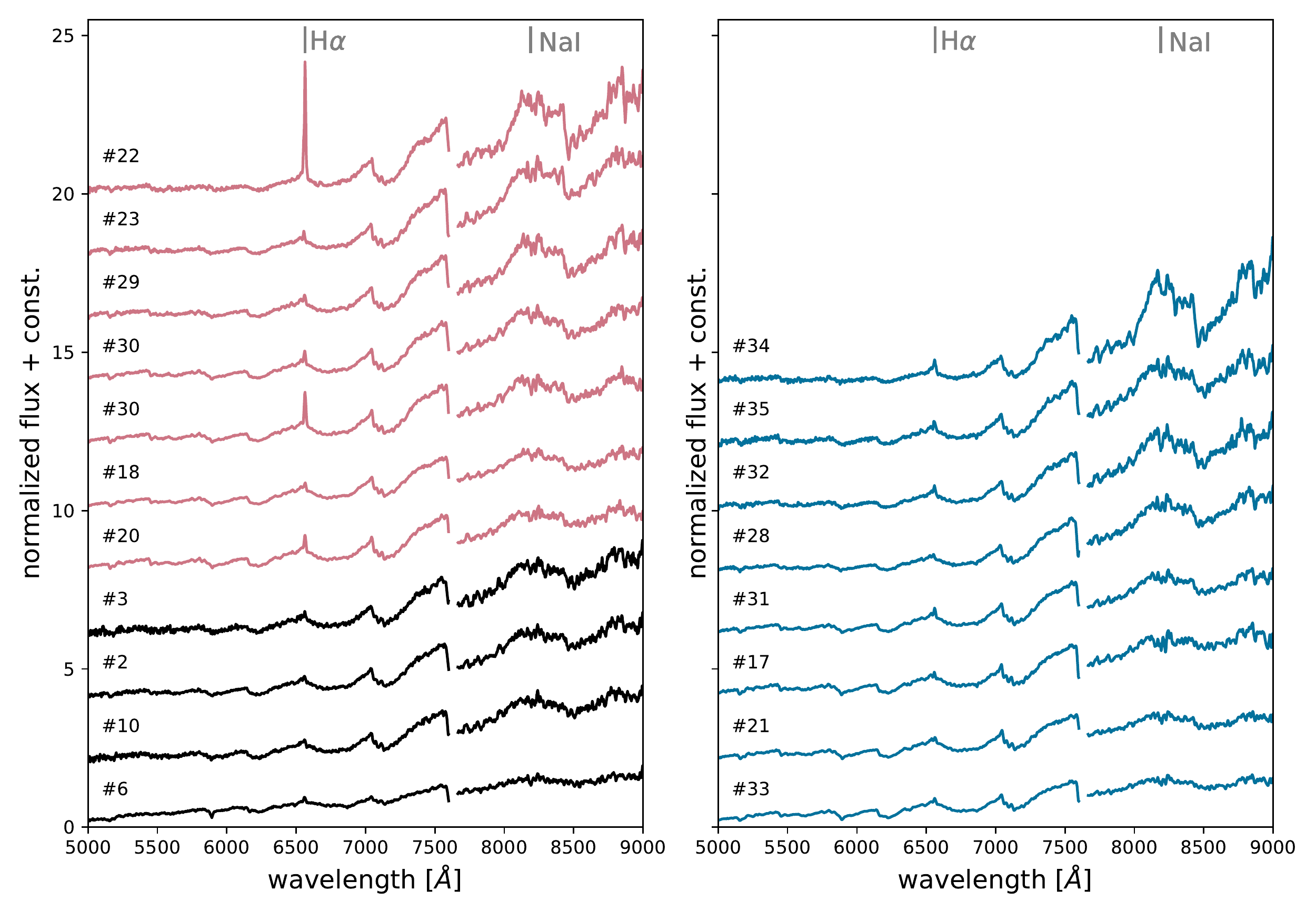}
      \caption{The spectra towards the Chamaeleon region taken with the FLOYDS spectrograph. The spectra towards region A shown in black, and those from region C in pink ($\delta < -77^{\circ}$) and blue ($\delta > -77^{\circ}$). Two spectra are shown for the object $\#30$, taken at different dates. A strong variability in H$_\alpha$ emission can be appreciated. The region around $\sim7600\,$\AA\ has been masked due to a pipeline artefact. The central wavelengths of H$_{\alpha}$ and Na\,I are indicated at the top of the figure.
              }
         \label{fig:spec_all}
   \end{figure*}
%-----------------------------------------------------------------

\subsection{Spectral type and extinction}
\label{sec:templates}
The spectra are shown in Fig.\,\ref{fig:spec_all} clearly show that all the observed objects are cool dwarfs of spectral type M, dominated by various molecular features (mainly TiO and VO; \citealt{kirkpatrick91,kirkpatrick95}). To derive more precise spectral types and assess if there is any extinction towards the objects, we compare the spectra to an empirical grid consisting of spectral templates:
%of field dwarfs, members of young star forming regions and the somewhat older TWA young moving group with spectral types M1 to M9.
%The analysis is similar to the one applied in the \citet{muzic14}, with a difference that here we do not include giant templates into the analysis, since Gaia DR2 parallax and magnitudes exclude this possibility.
\begin{itemize}
    \item \textit{field}:  a grid of field dwarfs with spectral types M1 to M9, separated by 0.5 spectral subtypes was created by averaging a number of available spectra at each sub-type\footnote{From \url{http://www.dwarfarchives.org, http://kellecruz.com/M_standards/}, and \citet{luhman03a, luhman04c}}.
  \vspace{0.1cm}
    \item \textit{young}: a grid of young objects (1-3 Myr) consists of the spectra of individual objects in Cha~I, Taurus, $\eta\,$Cha \citep{luhman03a, luhman04a, luhman04b, luhman04c}, and Collinder 69 \citep{bayo11} with spectral types M1 to M9, with a step of 0.25-0.5 spectral subtypes.

   \vspace{0.1cm}
    \item \textit{TWA}: spectra of members of the TW Hydrae association (TWA) from \citet{venuti19}, with spectral type range M3- M5.5 and M8-M9.5, with a step of 0.5-1 spectral subtypes. The age of TWA (8-10 Myr; e.g. \citealt{ducourant14, herczeg14, donaldson14, bell15}) make these spectra appropriate to compare with \epscha, expected to be of similar age.
\end{itemize}

The template spectra were smoothed and rebinned to match the resolution and the wavelength scale of the FLOYDS spectra, and the region around the H$_{\alpha}$ line was masked (strong H$_{\alpha}$ may be present both in our spectra and the young templates, which may affect the fit).
The spectral templates were reddened by A$_V$=0-5~mag, with a step of 0.2~mag, using the extinction law from \citet{cardelli89}.
The spectral fitting is performed over the wavelength range covered by the templates, typically 6100 - 9000 \AA.
The best fit spectral type and extinction were determined
by minimizing the reduced chi-squared value ($\chi^2$) defined as
\begin{equation}
 \chi^2=\frac{1}{N-m} \sum_{i=1}^{N} \frac{(O_i-T_i)^2}{\sigma_{O,i}^2},
\end{equation}
where O and $\sigma_O$ are the object spectrum and its uncertainty, respectively, T the template spectrum, N the number of data points, and m the number of fitted parameters (m=2).

\begin{figure}
   \centering
   \includegraphics[width=0.45\textwidth]{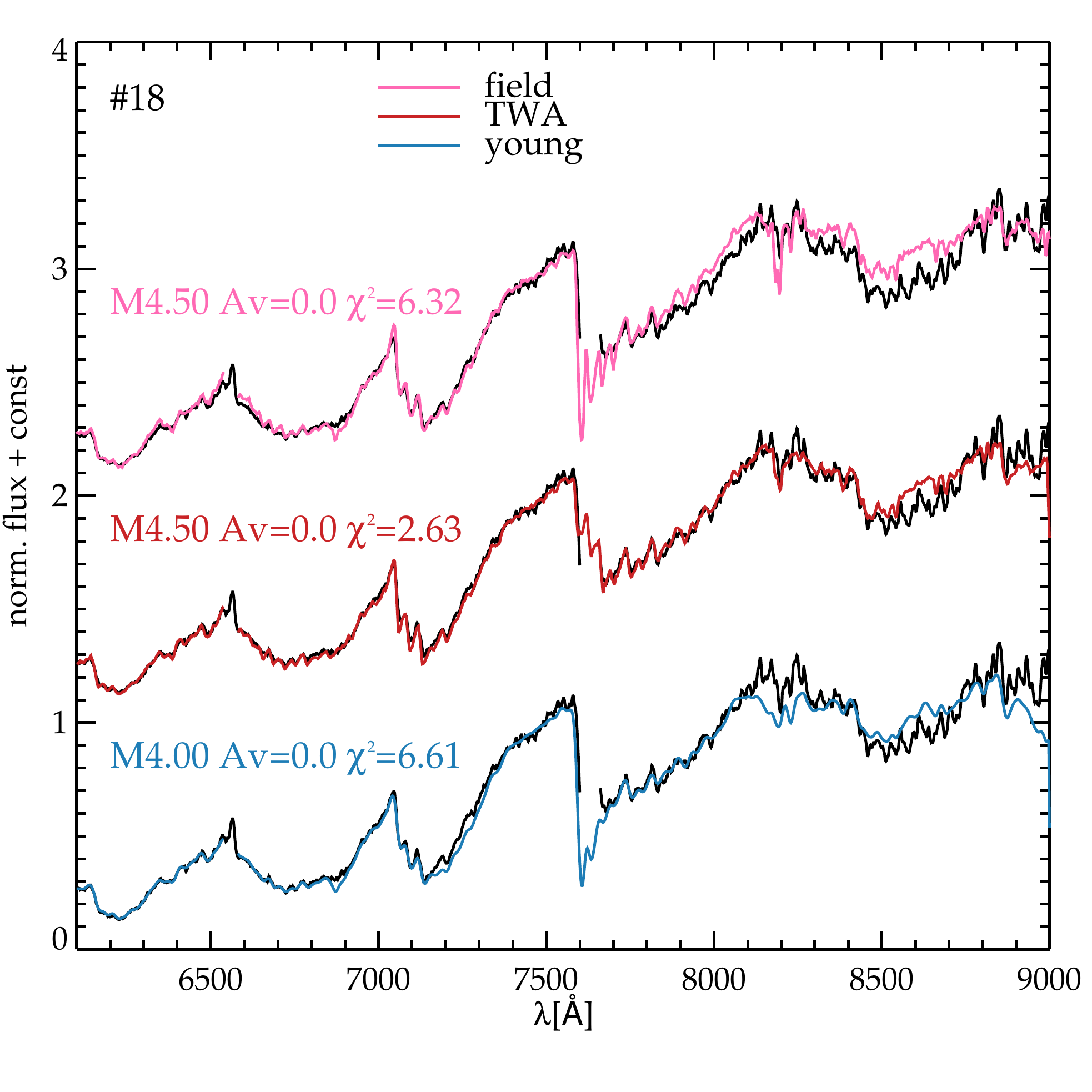}
      \caption{An example of the spectral fitting for the object $\#18$ (black). The best fit young (blue), TWA (red), and field (pink) template are shown, along with the corresponding spectral type, extinction, and the reduced $\chi^2$ of the fit. We see that the TWA template provides the best fit to the overall spectrum, as well as to the gravity sensitive sodium feature at $\sim8200$\,\AA.
              }
         \label{fig:spec_single}
   \end{figure}

An example of the best fit spectra for each set of templates is shown in Fig.\,\ref{fig:spec_single}, for the object $\#18$. The remaining fits are shown in Figs.~\ref{fig:spec_appendix1} and \ref{fig:spec_appendix2}.
Typically, several best-fit results cluster within $\pm1$ spectral subtype  or less, from the best-fit value, irrespective of the age of the objects in the template grid (field, TWA, or young). The extinction A$_V$ is zero, or very close to it, for all the objects.
Of the 18 objects, 16 are best-fit by either of the two younger categories (young or TWA), suggesting young age for most of the objects. This, however, is not enough to claim youth for these objects, and the final membership assessment requires a closer look into gravity-sensitive spectral features. The derived spectral types and extinctions are given in Table~\ref{tab:spt_par}.

%-----------------------------------------------------------------
%-----------------------------------------------------------------

% \include{table_spectral_parameters}
%When pasting the new version of the table, add:
%table to table*
%move caption to the top
%\hline \hline after begin{tabular}
%\hline before end{tabular}
%\begin{small} after begin{table*}
%\end{small} before end{table*}
%Below header:	 & (K) & (mag) & & (K) & (mag) &  & & (mag) & & (\AA) &  & \\

%{\color{blue} To check: eps Cha 05}}

\begin{table*}
\begin{small}
	\caption{Results of the spectral fitting, equivalent widths, and membership assessment.}
	\begin{center}
	\begin{tabular}{l l l l l l l l l l l l l}
	\hline \hline
			ID & T$_{\mathrm{eff}}^a$ & A$_{V}^a$ & log($g$)$^a$ & T$_{\mathrm{eff}}^b$ & A$_{V}^b$ & log($g$)$^b$ & SpT$^c$ & A$_{V}^c$ & best-fit templ.$^c$ & EW & youth sign.$^d$ & membership \\
			 & (K) & (mag) & & (K) & (mag) &  & & (mag) & & (\AA) &  & \\
			\hline
%	1 & 3000 & 1.0 & 4.0 &  &  &  &  &  &  &  &  X & Cha I \\ to ignore the WANG 2016 Xray
	1 & 3000 & 1.0 & 4.0 &  &  &  &  &  &  &  &   & Cha I \\
			2 & 3400 & 0.5 & 3.5 & 3500 & 2.20 & 3.5 & M4.00 & 0.2 & young &  -8.5 & NaI &  Cha I\\
			3 & 3000 & 0.0 & 5.0 & 3400 & 2.20 & 4.0 & M4.50 & 0 & TWA & -10.9 & NaI IRex &  Cha I \\
			4 & 2600 & 0.25 & 4.0 &  &  &  &  &  &  &  &  &  \\
			5 & 2700 & 0.25 & 5.0 &  &  &  &  &  &  &  &  IRex &  Cha I\\
			6 & 3600 & 0.25 & 4.0 & 3900 & 2.00 & 4.5 & M2.50 & 0 & field &  -4.9 &  & field \\
			7 & 3600 & 0.25 & 5.0 &  &  &  &  &  &  &  &  X & Cha I \\
			8 & 2800 & 0.25 & 4.5 &  &  &  &  &  &  &  &  &  \\
			9 & 3200 & 0.0 & 3.5 &  &  &  &  &  &  &  &  &  \\
			10 & 3400 & 0.5 & 4.0 & 3400 & 2.00 & 4.5 & M4.00 & 0 & young &  -7.0 & NaI & Cha I \\
			11 & 3400 & 1.0 & 4.0 &  &  &  &  &  &  &  &  &  \\
			12 & 4200 & 5 & 3.5 &  &  &  &  &  &  &  &  IRex &  Cha I\\
			13 & 3200 & 1.5 & 4.0 &  &  &  &  &  &  &  &  IRex &  Cha II\\
			14 & 3000 & 0.25 & 3.5 &  &  &  &  &  &  &  &  IRex &  Cha II\\
			15 & 2800 & 0.0 & 5.0 &  &  &  &  &  &  &  &  &  \\
			16 & 3300 & 0.0 & 3.5 &  &  &  &  &  &  &  &  &  \\
			17 & 3300 & 0.0 & 5.0 & 3500 & 1.40 & 4.5 & M4.50 & 0 & field &  -4.9 &  & field \\
			18 & 3200 & 0.25 & 4.0 & 3500 & 1.00 & 3.5 & M4.50 & 0 & TWA &  -9.3 & NaI &  \epscha\\
			19 & 3200 & 0.25 & 4.0 &  &  &  &  &  &  &  &  X &  \epscha\\
			20 & 3100 & 0.0 & 4.0 & 3500 & 1.00 & 3.5 & M4.50 & 0 & TWA & -15.5 & NaI & \epscha \\
			21 & 3400 & 0.25 & 4.0 & 3500 & 0.60 & 4.0 & M4.50 & 0 & TWA &  -6.3 & NaI & \epscha \\
			22 & 2800 & 0.25 & 4.0 & 3300 & 2.20 & 3.5 & M5.75 & 0 & young & -123.7 & H$_\alpha$ NaI IRex &\epscha  \\
			23 & 3000 & 0.25 & 4.0 & 3300 & 2.40 & 4.0 & M5.75 & 0 & young & -12.1 & NaI & \epscha \\
			24 & 3500 & 0.0 & 4.0 &  &  &  &  &  &  &  &  X &  \epscha\\
			25 & 2500 & 0.0 & 4.5 &  &  &  &  &  &  &  &  &  \\
			26 & 2800 & 0.0 & 4.5 &  &  &  &  &  &  &  &  &  \\
			27 & 3200 & 0.0 & 5.0 &  &  &  &  &  &  &  &  &  \\
			28 & 3200 & 0.25 & 4.0 & 3400 & 2.20 & 4.0 & M4.50 & 0 & TWA &  -7.5 & NaI & \epscha \\
			29 & 3000 & 0.0 & 5.0 & 3400 & 1.60 & 3.5 & M5.50 & 0 & TWA & -14.0 & NaI & \epscha \\
			30 & 3200 & 0.25 & 4.0 & 3400 & 1.00 & 4.0 & M4.50 & 0 & young & -30.4, -14.7 & H$_\alpha$ NaI & \epscha \\
			31 & 3200 & 0.25 & 4.0 & 3400 & 1.00 & 4.0 & M4.50 & 0 & TWA & -10.9 &  NaI & \epscha \\
			32 & 2900 & 0.0 & 5.0 & 3200 & 1.80 & 4.5 & M4.50 & 0 & TWA & -15.2 &  IRex & \epscha \\
			33 & 3400 & 0.0 & 5.0 & 3600 & 1.00 & 4.5 & M3.00 & 0 & TWA &  -5.6 & NaI & \epscha \\
			34 & 2700 & 0.25 & 4.5 & 3000 & 2.80 & 4.0 & M7.50 & 0 & young & -29.2 & NaI IRex & \epscha \\
			35 & 3000 & 0.0 & 5.0 & 3300 & 2.20 & 4.0 & M4.50 & 0 & TWA & -14.7 & NaI IRex & \epscha \\
			\hline
		\end{tabular}
		\tablefoot{$^{a}$ from the SED fitting; $^{b}$ from the atmosphere model fitting; $^{c}$ from the spectral template fitting; $^{d}$ X: X-ray source; NaI: shallow Na~I doublet; IRex: exhibiting infrared excess; H$\alpha$: H$\alpha$ compatible with accretion.
		X-ray detections are from \citet{alcala95}, classified as WTTS therein.
	}
	\end{center}
	\label{tab:spt_par}
	\end{small}
\end{table*}

\subsection{Indicators of youth and membership}

    \begin{figure}
   \centering
   \includegraphics[width=0.48\textwidth]{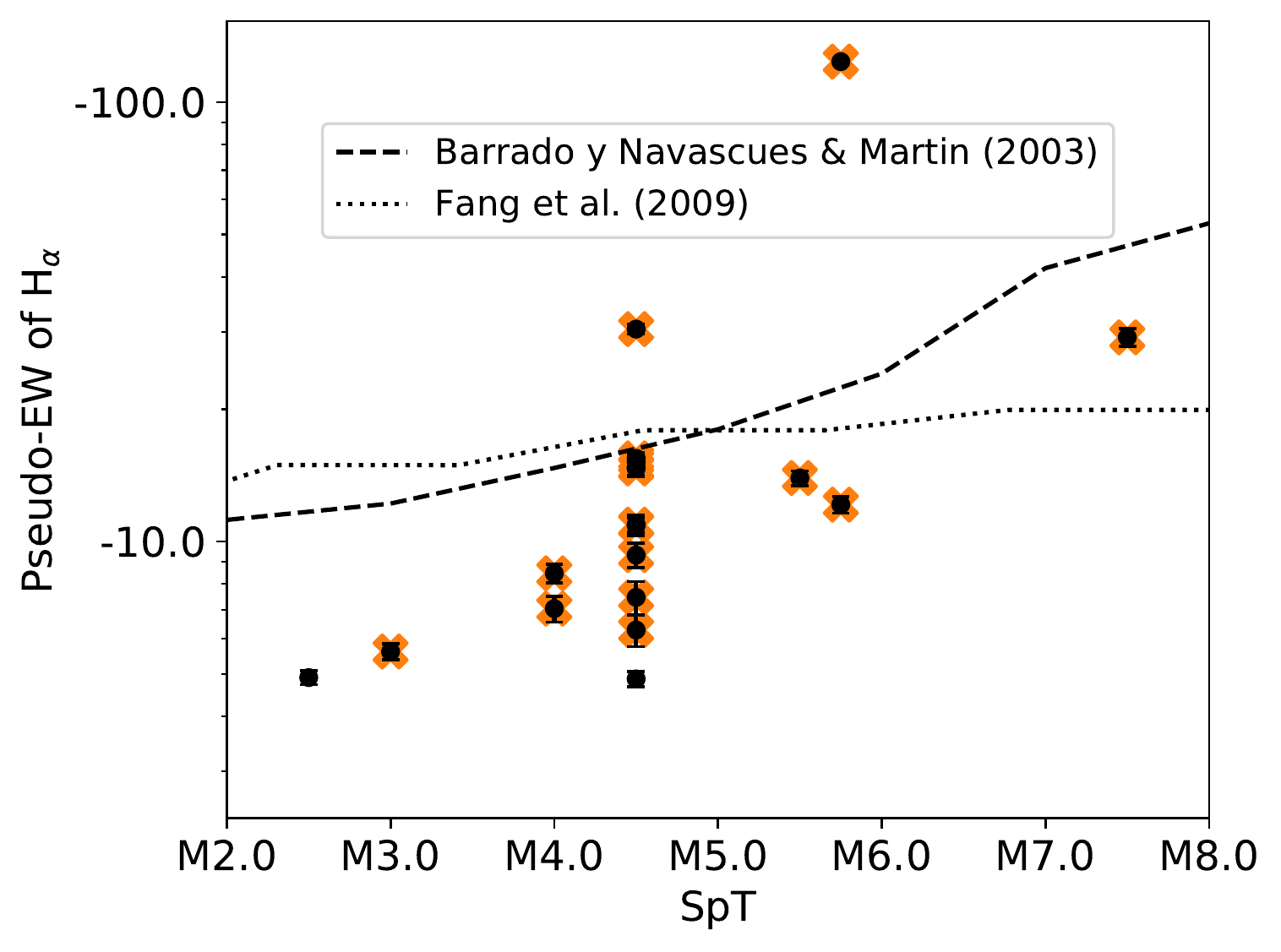}
      \caption{Pseudo-EW measurement for the candidates with spectra (black dots). Two measurements for the object $\#30$, separated by about a month, are additionally marked with open circles, and the objects with other signatures of youth (Na I doublet, X-ray detection or IR-excess) with orange crosses. The criteria separating the accretors from the non-accreting sources are shown by a dashed \citep{barrado03} and dotted \citep{fang09} lines.}
         \label{fig:ew}
   \end{figure}

As demonstrated in Fig. 6 of \citet{muzic14}, optical spectra of M-dwarfs provide several gravity-sensitive features that allow us to distinguish young from the field stars, even at very low spectral resolution of a few hundred. The two main features that we will rely on in this analysis are (i) the Na\,I doublet at 8183/8195\,\AA, which becomes more pronounced (i.e. larger equivalent widths, and deeper absorption) at high surface gravities (e.g. \citealt{martin99, riddick07,schlieder12}),
and (ii) the H$_{\alpha}$ emission, which is a sign of accretion, and therefore youth.

H$_{\alpha}$ emission is commonly present in the spectra of M-type main sequence stars as a result of a chromospheric activity, whereas in young stars both the activity and accretion may contribute. One way to identify the accreting sources is by means of elevated equivalent widths (EWs) of H$_{\alpha}$ with respect to non-accreting pre-main sequence and main-sequence stars \citep{white&basri03,barrado03,fang09,alonso15}. We measure the pseudo-EWs\footnote{EW with respect to a local pseudo-continuum, formed by molecular band absorptions dominating the optical spectra of cool dwarfs}. The results are shown in Fig.\,\ref{fig:ew} as a function of the spectral type. The error bars are calculated as standard deviations from repeated measurements obtained by slightly changing the wavelength range for the pseudo-continuum subtraction, and the range over which the EW is measured.

The criteria used to distinguish between accretion and activity from \citet{barrado03} and \citet{fang09} are overplotted. Two sources are located above  both lines ($\#22$ and $\#30$), while the source $\#34$ "passes" only the Fang et al. criterion. We note, however, that the equivalent plot from that work is more sparsely populated for spectral types later than M6 that that of \citet{barrado03}. The object $\#30$ has been observed twice, with about a month passing between the observations. The two measurements are highlighted with open circles in Figue~\ref{fig:ew}, one of them falling just below the accretion criterion. Both continuum and emission line variability commonly occur in low-mass pre-main sequence stars \citep[e.g.][]{nguyen09}. The H$_{\alpha}$ variability in several $\sigma$ Ori ($\sim$3 Myr) low-mass stars has been interpreted as consistent with the presence of cools spots with the filling factors between 0.2 - 0.4 \citep{bozhinova16}. Three more sources at the same spectral type as $\#30$ (M4.5) fall close to it on its lower accretion levels ($\#20$, $\#32$, $\#35$).

The strongest accreting object in our sample, candidate $\#22$, has recently be reported by \citet{schutte20}, who identify the object as a member of \epscha, showing a strong infrared excess in the WISE data. The spectral type measured from the near-infrared spectra, M6$\pm$0.5, is in agreement with the spectral type found here.

We closely inspected all the spectra to identify the presence or absence of H$_\alpha$ emission, and the strength of the Na\,I absorption. A dedicated column in Table~\ref{tab:spt_par} signals the presence of H$_\alpha$, and a shallow Na\,I absorption, which we take as signatures of the young age of the source. The Na\,I criterion is a stronger one when assigning the membership to the sources, as not all young stars may be accreting.
Furthermore, the excess at mid-infrared wavelengths, signalling a presence of a disk or an envelope, as well as strong X-ray emission are both well-established signatures of youth. \cite{alcala95}  {\new  classified these X-ray emitters as WTTS based on presence of Li I ($\lambda$6707\AA) absorption line and/or  H$_{\alpha}$ emission line.} The objects showing either of the two properties have been additionally labelled  with "IRex" and/or "X" in Table\,\ref{tab:spt_par}.  Objects with one or more signatures of youth are considered members of the corresponding star forming region or young moving group.
%While the age category assignment is straightforward for most of the objects in our sample, one case appears more uncertain. The object \#32 is best fit by a TWA-class template but the pronounced Na\,I absorption resembles better the spectra of the field objects. We therefore consider the membership of this object uncertain, as the weak H$_\alpha$ emission may also have a chromospheric origin \citep[e.g.,][]{Walkowicz08}.

\subsection{Comments on specific objects}
A few of the objects on our candidate list, while not appearing on the member lists considered in Sec.~\ref{sec:members}, have been studied in some detail in other works. We have searched VizieR \citep{vizier} and Simbad \citep{simbad} databases for additional information on the candidates.

\begin{itemize}
    \item
The object $\#7$ seems to be a known X-ray-emitting spectroscopic binary, and a member of Cha I. Known also as RX J1108.8-7519A, it appears on members lists of \citet{luhman08} and \citet{esplin17} with a spectral type K6, and has been classified as WTTS by \citet{wahhaj10}, and Class III candidate by the machine learning classifier based on WISE photometry \citep{marton16}.

\item
The object $\#12$ has previously been identified as a Class I/II YSO candidate by \citet{marton16}.

\item
The object $\#13$ (AY Cha) has been listed as a RR Lyrae variable in the General Catalogue of Variable Stars \citep{GCVS}. It has also been identified as a Class I/II YSO candidate by \citet{marton16}, consistent with the SED shown in Fig.\,\ref{fig:sed} which clearly shown an infrared excess. Knowing that RR Lyrae are stars at a late stage of their evolution, typically associated with globular clusters and high galactic latitudes, we find this classification highly unlikely. We inspected the lightcurve from the Transiting Exoplanet Survey Satellite (TESS; \citealt{TESS}) and find that the source shows a clear sinusoidal periodicity, and not the characteristic RR Lyrae-type shape. We conclude that it is likely that this object is a young pre-main sequence star.

\item The objects $\#$19, 26, and 27 have been included in the list of provisional members of \epscha~ in \citet{murphy13}, pending further confirmation. The object $\#$27 was classified as member by \citet{feigelson03} with the label \epscha 12, but discarded by \citet{murphy13} as an outlier based on the proper motion measurements. The new astrometric information provided by \Gaia\ speaks in favour of their membership, although none of these three objects were part of our spectroscopic follow-up.

\end{itemize}

\subsection{Comparison to atmospheric models}
 \begin{figure*}
   \centering
   \includegraphics[width=0.95\textwidth]{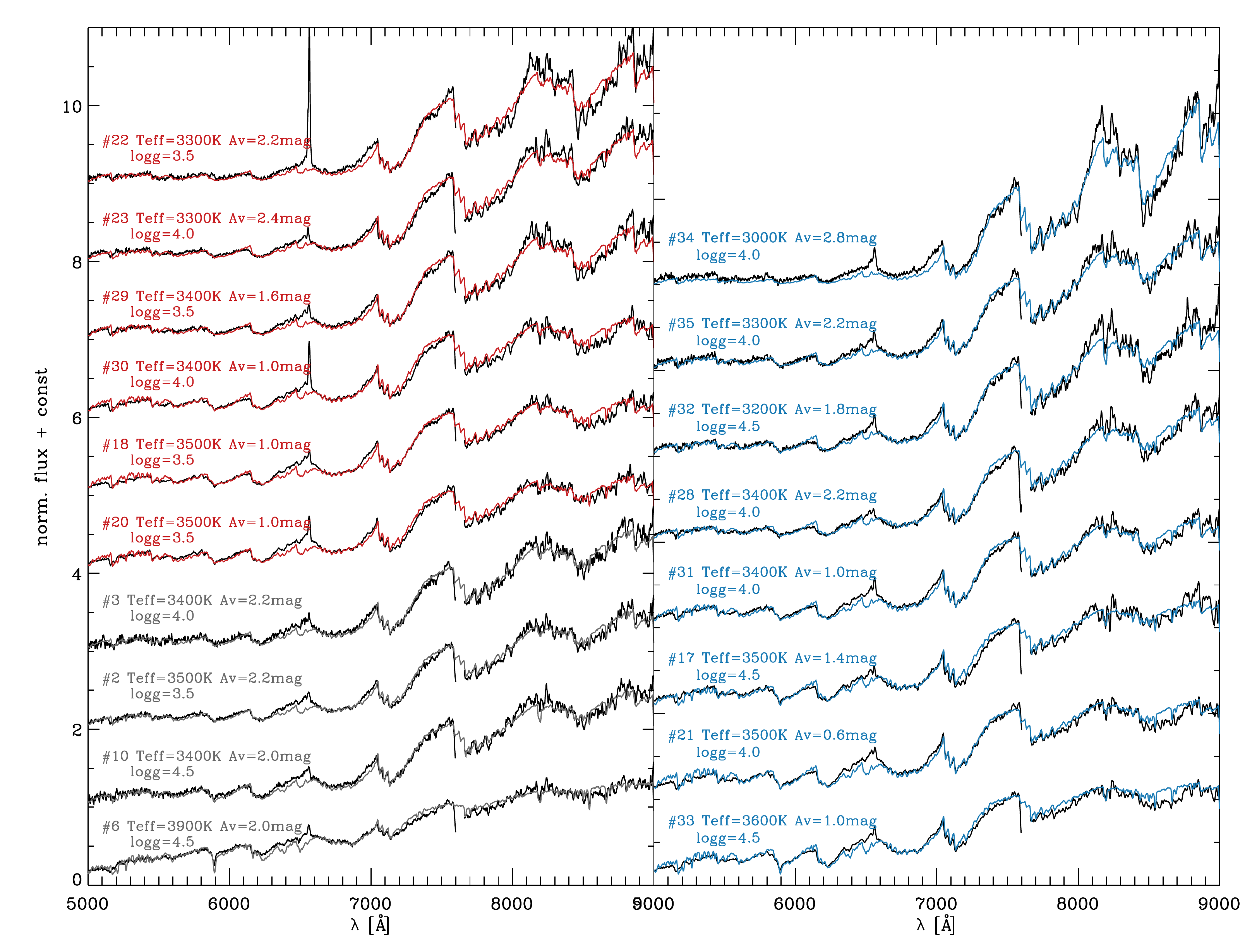}
      \caption{The spectra of our candidates, along with the best-fit
        BT-Settl model. The models are colour-coded depending on the
        membership candidacy (grey for Cha I, red \epscha, and blue
        \epscha/LCC).
      }
      \label{fig:models}
 \end{figure*}

In this section we use the the FLOYDS spectra, and apply $\chi^2$
minimisation to find the best fit atmosphere model, and its
corresponding $T_{\mathrm{eff}}$, extinction A$_V$, and log$g$. For
this, we use the BT-Settl models \citep{allard11} in the
$T_{\mathrm{eff}}$ range between 2000 and 5000 K, in steps of 100 K
and log$g$=3.5 - 5.0 in steps of 0.5 dex. The A$_V$ is varied between
0 and 5\,mag, with the step of 0.2\,mag. As before, the extinction law
from \citet{cardelli89} is employed to redden the models prior to
$\chi^2$ calculation. The spectral fitting is performed over the full
wavelength range of our spectra, from 5000 to 9000 \AA. The best-fit
parameters for each object are given in Table~\ref{tab:spt_par}, and
shown in Fig.\,\ref{fig:models}. {\new The fitting procedure is the
  same as in Section~\ref{sec:SED}  but here performed on the optical
  spectra instead of the broad band photometry. In both cases the selected models and
  parameter ranges are the same. In Section~\ref{sec:SED} the SED
  fitting is done for all selected candidates in Cha I, Cha II and
  \epscha\ , here we fit the parameters only for 18 sources with FLOYDS
spectra.}

While the overall shape of the spectra can be reasonably well
represented by the models, the region around the H$_\alpha$ line does
not match the shape of the spectra well. This region was anyway
excluded from the fitting procedure in order to avoid discrepancies
due to H$_\alpha$ emission present in some of the candidate
objects. The best-fit models show preference towards low surface
gravity: {\newst{14/18}} {\new 13/18} objects have log$g$=3.5 - 4, which is characteristic
for young low-mass stars and brown dwarfs. The remaining {\newst{4}}
{\new 5} objects have the best-fit log$g$=4.5. The $T_{\mathrm{eff}}$ and A$_V$ obtained by atmosphere model fitting tend to be systematically higher than those obtained both from the SED fitting ($\Delta T_{\mathrm{eff}}$=260\,K, $\Delta$A$_V$=1.5\,mag on average), and the template fits ($\Delta$A$_V$= 1.7\,mag). Most of our candidates are located in off-cloud regions, i.e. the extinction close to zero may be expected. A similar trend is present in the results of \citet{bayo17}, when comparing the $T_{\mathrm{eff}}$ and A$_V$ derived from fitting the BT-Settl models to the optical spectra, and the SED fitting. On the other hand, they also report
a trend towards determination of higher temperatures when using only optical data versus including the near-infrared spectra in the fit. In the latter case, the
best-fit parameters are more in line with those from the SED fitting, which is probably not surprising given that the SEDs in \citet{bayo17}, as well as in this work, contain both optical and near-infrared photometry.

\section{Discussion}

\subsection{Relation between \epscha~and LCC}
\label{sec:regionC}

\begin{figure*}[htb]
    \centering
    \includegraphics[width=0.9\linewidth]{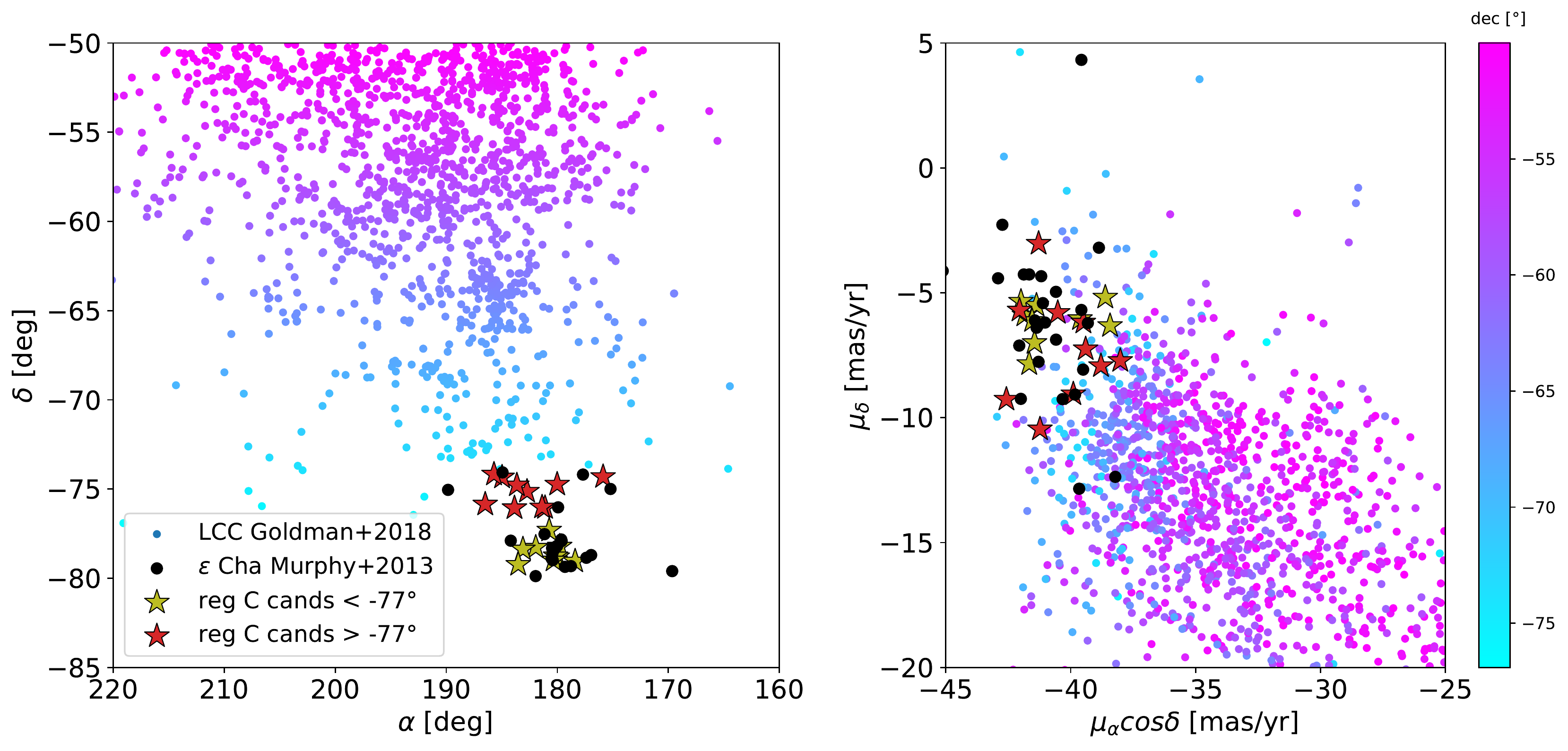}
    \includegraphics[width=0.9\linewidth]{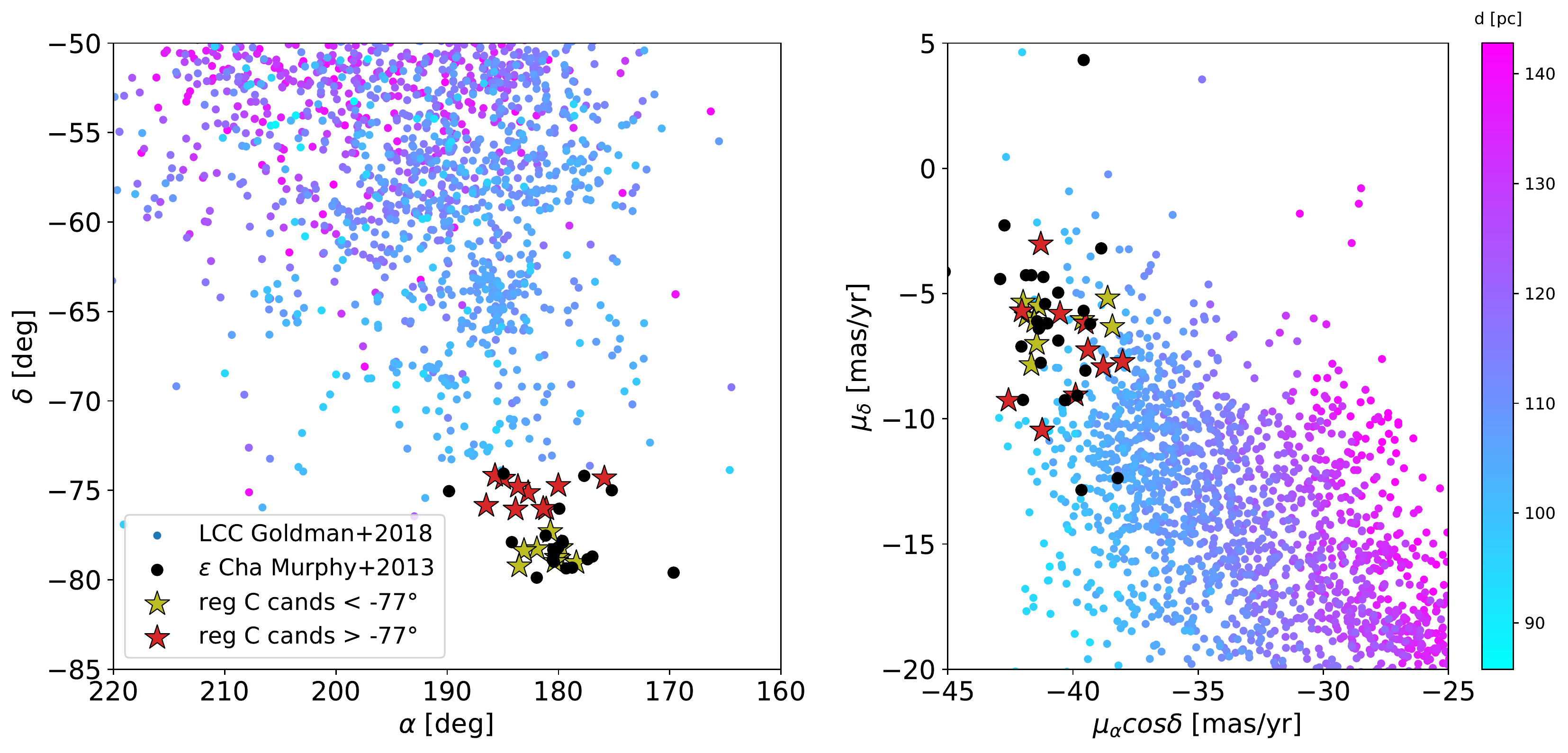}
    \caption{On-sky positions and proper motions of the candidates from region C (red and green stars), along with the members \epscha~ (black circles \citealt{murphy13}), and Gaia DR2-selected candidate members of LCC (colored dots \citealt{goldman18}). LCC sources are coloured according to the declination (top plots), and Gaia DR2 distance (bottom). The distances of \epscha~ objects in this plot are in the range $\sim$ 93 - 110 pc, similar to our candidates ($\sim$ 94 - 113 pc). }
    \label{fig:regionIII}
\end{figure*}

Candidates from the region C overlap both with the proposed members of \epscha~\citep{murphy13}, and, towards the north, with the kinematic candidate members of the LCC A0 subgroup \citep{goldman18}. Moreover, there are several sources in common between the two lists. As already pointed out by \citet{murphy13}, given the similarity in ages, kinematics, distances and the observed trends in these parameters, a useful demarcation between the southern extent of LCC and \epscha~may simply not exist. In this section we try to shed some light on the relation between these two regions, based on the improved astrometry provided by Gaia DR2.

As evident from the surface density plots (Fig.\,\ref{fig:regions}), there appears to be a gap between the region where the main bulk of \epscha~members reside, and the northern portion where the LCC candidates overlap with a smaller portion of the \epscha~objects. To investigate the properties of our region C candidates associated with the two overdensities observed in Fig.\,\ref{fig:regions}, we divide them according to the declination, as those south or north of $\delta=77^{\circ}$. They are plotted as red and green stars in Fig.\,\ref{fig:regionIII}, where we show their on-sky positions and \Gaia\ DR2 proper motions, along with the members of \epscha~(black dots), and LCC candidates. The latter are colour-coded according to declination (top plots), and distance (bottom). The LCC candidates show a clear gradient in proper motions both with the on-sky position, as well as with the distance. Both \epscha~members, and our objects share the proper motion space with the southern-most subgroup of LCC (A0), and cannot be separated neither according to the proper motion, nor the distance. The mean values and the standard deviations of these quantities are:
\begin{itemize}
    \item Candidates with $\delta<-77^{\circ}$ (southern portion):\\
    $\mu_{\alpha}$cos$\delta$ = $-40.73 \pm 1.35\,$mas\,yr$^{-1}$,\\
    $\mu_{\delta}$ = $-6.14 \pm 0.80\,$mas\,yr$^{-1}$,\\
    $\varpi=9.58 \pm 0.29\,$mas;
    \item Candidates with $\delta>-77^{\circ}$ (northern portion):\\
    $\mu_{\alpha}$cos$\delta$ = $-40.32 \pm 1.39\,$mas\,yr$^{-1}$,\\
    $\mu_{\delta}$ = $-7.24 \pm 2.04\,$mas\,yr$^{-1}$,\\ $\varpi=9.80 \pm 0.39\,$mas;
\end{itemize}

\begin{figure*}
   \centering
   \includegraphics[width=0.8\textwidth]{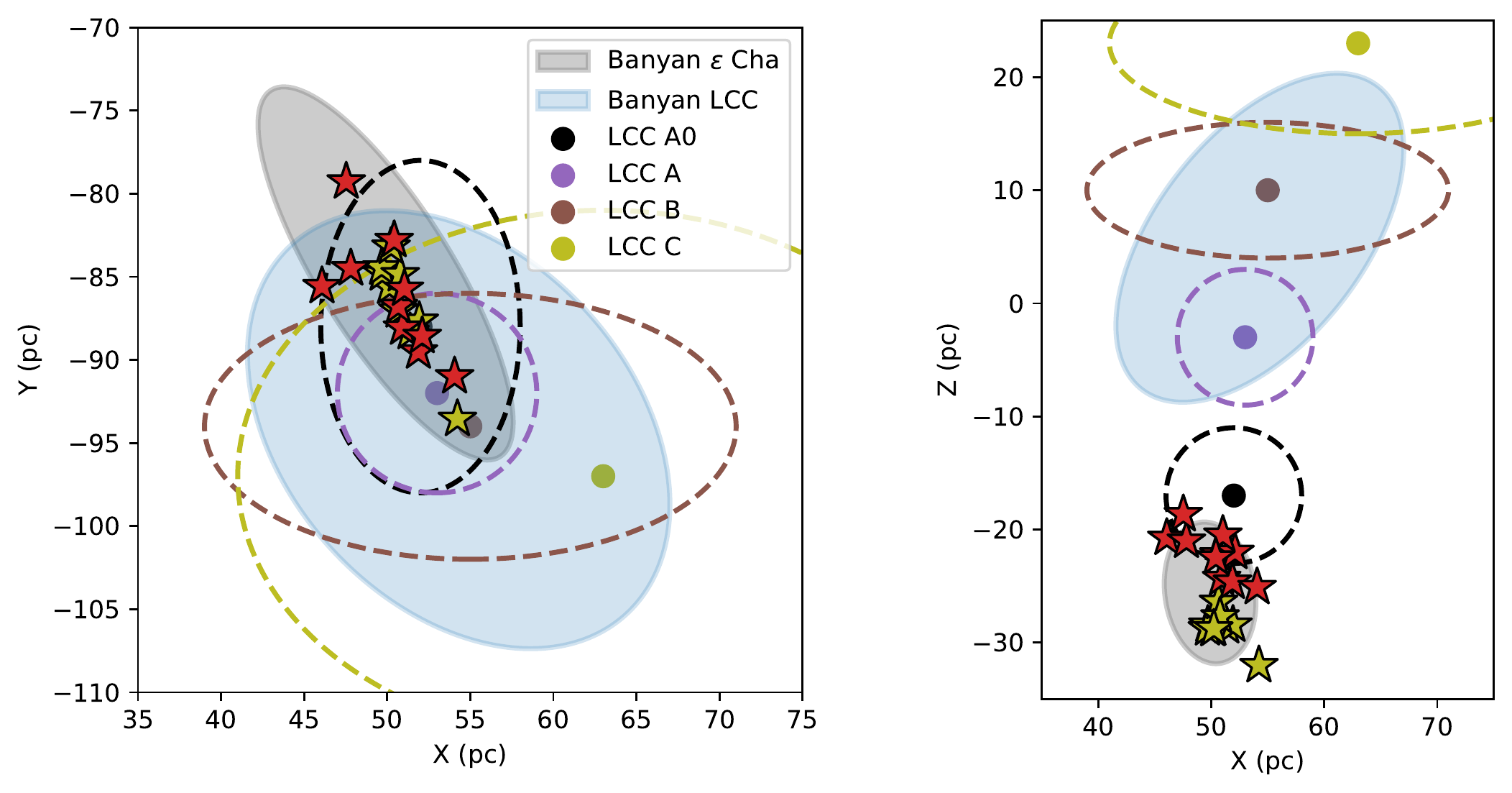}
      \caption{Distribution of the region C candidates in the Galactic XYZ coordinates, where the positive X axis points towards the Galactic Centre, Y is in the direction of the Galactic rotation, and Z points towards the North Galactic pole. The candidates are represented by stars, with the colour-coding identical to that of Fig.\,\ref{fig:regionIII}.
      The shaded ellipses mark the areas occupied by \epscha~and LCC members from the BANYAN $\Sigma$ model \citep{gagne18}. The dots mark the mean positions of the LCC subgroups and the semi-major/minor axes of the dashed ellipses equal 2 standard deviations   \citep{goldman18}.
              }
         \label{fig:XYZ}
   \end{figure*}

To try to understand better the relation of our candidates with the two structures as per their positions in the Galaxy, in Fig.\,\ref{fig:XYZ}, we show their distribution in the Galactic XYZ coordinate frame, with the axes pointing towards the Galactic Centre (X), in the direction of the Galactic rotation (Y), and towards the North Galactic pole (Z). Furthermore, we plot the space occupied by \epscha~and LCC according to the BANYAN kinematic models of the nearby moving groups (shaded ellipses; \citealt{gagne18}), and the LCC subgroups A0, A, B and C from \citep{goldman18}. While in the XY plane (left panel in Fig.\,\ref{fig:XYZ}) all of the structures show a significant overlap, there is a clear distinction between several structures perpendicular to the Galactic plane. The BANYAN definition of LCC overlaps with the Goldman groups A, B, and C, whereas the A0 group appears as a bridge between those groups and \epscha. According to the BANYAN $\Sigma$ tool \citep{gagne18}, all our candidates have a high probability of being members of \epscha, and mostly non-existent one belonging to the LCC, the reason being evident from Fig.\,\ref{fig:XYZ}. The candidate $\#$17 has an 87.5\% probability of belonging to \epscha, 8.2\% to LCC, and 4.3\% to the field. All the remaining candidates have $\gtrsim 97\%$ probability of belonging to \epscha.The Bayesian probabilities reported by BANYAN $\Sigma$ are designed to generate recovery rates of 82\% when proper motion and distance are used, and 90\% when proper motion, radial velocity and distance are used. Of the 19 region C candidates, 3 have radial velocity reported in the literature.
The age of \epscha~derived by \citet{murphy13} is 3.7$^{+4.6}_{-1.4}$ Myrs, whereas Goldman et al. show the age distribution of A0 that peaks at $\sim$ 7 Myr, but extends from $\sim$ 3 Myr to $>$10 Myr. Given both the statistical uncertainties of these estimates, as well as the systematics associated with age derivation from the HR diagrams (uncertainties in distances and bolometric corrections, and use of different sets of evolutionary models in the two works), these two estimates are basically indistinguishable. The same is true for the two spatial subgroups of region C defined by the dividing line $\delta=-77^{\circ}$, where the northern subgroup has the mean age determined from the HR diagram (Fig.\,\ref{fig:hrd}) of $7\pm5$\,Myr, and the southern one   $6\pm3$\,Myr.

To conclude, given the present evidence on the age, parallax, proper motions, and the position in the Galaxy, it cannot be excluded that \epscha~and the southern extension of the LCC (A0) may have been born during the same star formation event.

\subsection{HR diagram}
\begin{figure}
   \centering
   \includegraphics[width=0.45\textwidth]{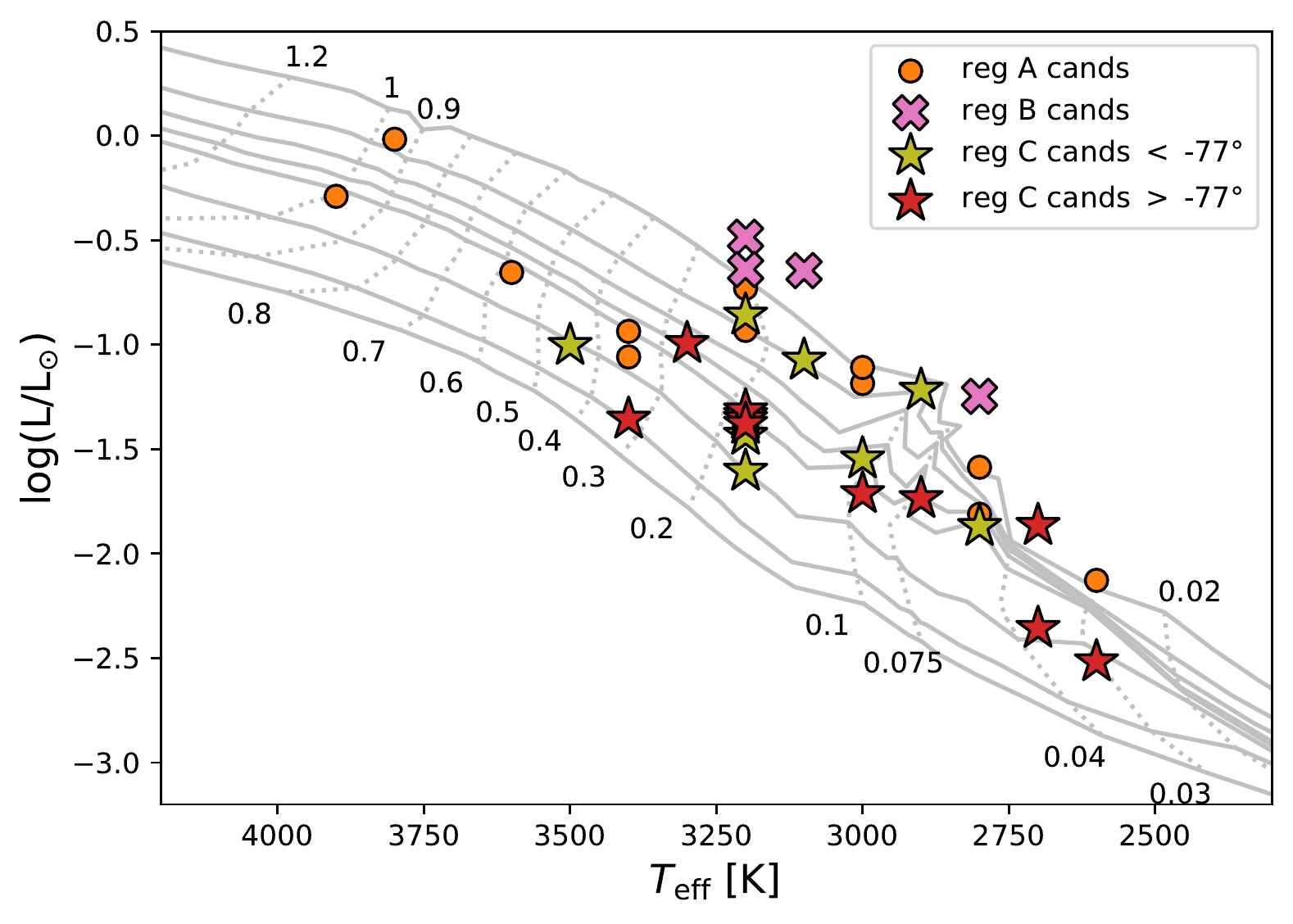}
      \caption{HR diagram for the candidates, along with the BT-Settl isochrones (solid lines) and the lines of constant mass (dotted lines). The isochrones correspond to the ages (1, 2, 3, 4, 5 (darker line), 10, 20, 30) Myr, from top to bottom, and the masses are given in M$_\odot$. The uncertainties in $T_{\mathrm{eff}}$ are 100 K, reflecting the spacing of the grid, while those in log$L$ are comparable to or smaller than the symbol sizes.
              }
         \label{fig:hrd}
   \end{figure}

To construct the Hertzsprung-Russell (HR) diagram shown in Fig.\,\ref{fig:hrd}, we use the $T_{\mathrm{eff}}$ and log$L$ determined by VOSA SED fitting. The bolometric luminosity is calculated from the total observed flux, which VOSA determines by integrating the best-fit model, multiplied by the dilution factor (R/d)$^2$, where R is the radius estimated in the fitting process, and d is the distance. All 35 candidates are shown in Fig.\,\ref{fig:hrd}, along with the isochrones (1-30 Myr) and the lines of constant mass from the BT-Settl series. The candidates belonging to region B (Cha II) appear the youngest of the whole sample ($\lesssim$1\,Myr), which is corroborated by the IR excess exhibited by 2 of them.  The Region A (Cha I) candidates are mostly consistent with ages $\lesssim$5\,Myr, with a mean age of $3\pm2$\,Myr.
The Region C is the oldest, with a mean age of $6\pm4$\,Myr. This is consistent with the results from fitting spectral templates. In most cases the best-fitting spectral template is from objects in the TW Hya association, indicating that \epscha\ has a similar age. There is no significant age difference between candidates from the northern (age of $7\pm5$\,Myr) and southern (age of $6\pm3$\,Myr) part of the region.

\subsection{Updated census in \epscha~and the IMF}
\cite{murphy13} compiled a list of 35 members of \epscha~ based on common kinematics.
Our selection yields additional 18 candidates, of which 14 have been included in the spectroscopic follow-up, confirming 13 of them as young sources. Additionally, sources $\#$19 and $\#$24 are X-ray emitters, which indicates their youth. Our study increases the census of confirmed members of \epscha\ by 15 new members, which is 43\% more. Out of 14 sources for which we obtained spectra, only one seems to belong to the field (see Fig. \ref{fig:spec_appendix2}), i.e. the confirmation rate of our selection method is larger than 90\%. For this reason, we included all of our candidates except the one rejected source in the sample used to calculate the Initial Mass Function (IMF). The final sample for the IMF therefore contains the previously known, confirmed, and candidate members, which is in total 52 objects.

The spectral types obtained from spectral fitting indicate that only one member of \epscha\ (source $\#$34) is a brown dwarf, it has the latest spectral type of M7.5. All except one member have spectral types later than M4.5. Our spectral fitting was done with steps of 0.5 -- 1 sub-types (see Sec.\,\ref{sec:templates}). Recently source $\#$20 has been reported by \citet{schutte20}, who identify the object as accreting brown dwarf with spectral type of M6 $\pm$ 0.5 based on infrared spectra. Our fit to optical templates results in spectral type M4.5 $\pm$ 1. Taking this into account together with the positions of all sources in the HR diagram, our selection contains: 3 brown dwarfs in Cha I and 5 brown dwarfs in \epscha. The IMF estimation for \epscha\  presented below is to the best of our knowledge the first time reaching the sub-stellar mass regime for this young moving group.

We obtained a "mass distribution" for each of the objects by running VOSA on 100 different photometry sets. In each realisation the photometry is moved by a random offset within the respective photometry errors (assuming Gaussian distributions). It is worth to mention that VOSA did not produce results (mass and/or age) in some runs, if the source lies outside the area covered by the isochrones in HR diagram. In order to obtain an evenly sampled mass distribution, we run a Monte Carlo simulation for each of the 52 sources. The Monte Carlo simulation is very similar to the one previously applied by \cite{muzic19}. The mass of each star is drawn from the distribution derived by multiple realisations of VOSA. This is performed 100 times, and for each of the 100 realisations we do 100 bootstraps, i.e., random samplings with replacement. In other words, starting from a sample with $N\leq100$ members, in each bootstrap we draw a new sample of 100 members, allowing some members of the initial sample to be drawn multiple times. This results in $10^4$ samples of the mass distributions for each star, which are used to derive ${\rm d}N/ {\rm d}M$ and the corresponding uncertainties.

\begin{figure}
   \centering
   \includegraphics[width=0.45\textwidth]{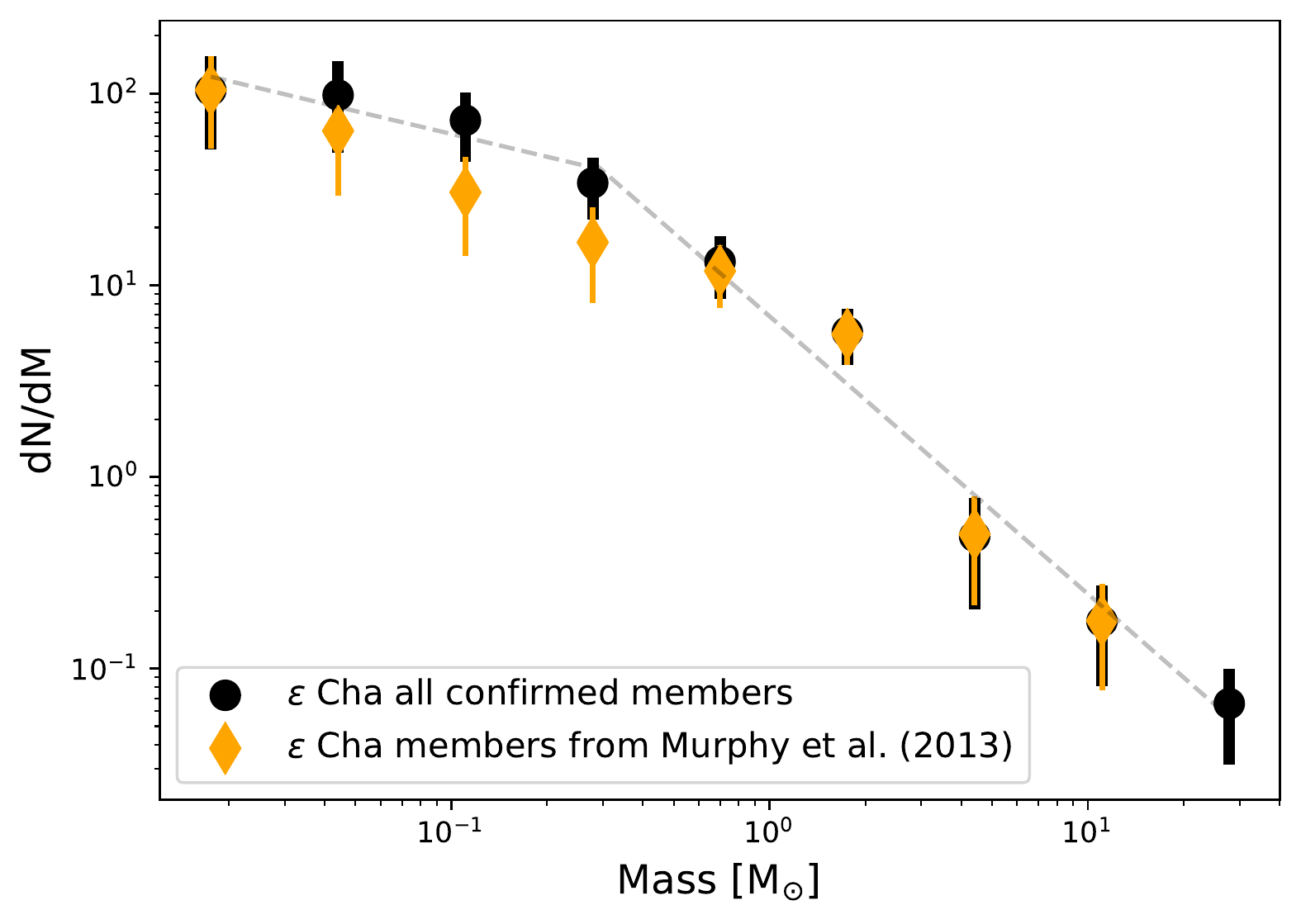}
      \caption{IMF of the \epscha~updated census, represented in equal-size bins of 0.2dex in mass. We plot the results from Monte Carlo simulations of mass distribution and corresponding uncertainties. In black the distribution from updated census, in yellow the distribution based on \cite{murphy13} our membership.
              }
         \label{fig:imf}
   \end{figure}

In Fig.\ref{fig:imf}, we show the IMF and the corresponding fits in the power-law form ${\rm d}N/{\rm d}M\propto M^{-\alpha}$. The IMF can be represented by two power laws with a break around 0.5 M$_{\odot}$. For M $<$ 0.5 M$_{\odot}\,\, \alpha = 0.42 \pm 0.11$ and for M $>$ 0.5 M$_{\odot} \,\,\alpha = 1.44 \pm 0.12$. Both values are in agreement with the power-law slope of low-mass IMF in various star forming regions, compare Table 4 in \cite{muzic17}. This IMF estimation is only illustrative since the membership of \epscha~ is still not complete, especially in the sub-stellar regime.

We estimate the completeness of the spectroscopically  observed sample in \epscha\ to be $G_{BP}=18.5$\,mag. Photometry reaches fainter stars when compared to spectroscopy. We estimate the completeness of photometric data used in our selection to be two magnitudes deeper, $G_{BP}=20.5$\,mag. This corresponds to a mass of $M= 0.02-0.03$ M$_{\odot}$ for an age of 5 Myrs.

Additionally we do not take into account any information about multiplicity of the objects, even though the majority of the members selected in \cite{murphy13} are known to be spectroscopic binaries.

\subsection{Distance to \epscha~}

The distance to \epscha~ was estimated based on \Gaia\ DR2 data, in a way similar to \citet{muzic19}.  We used a maximum likelihood procedure  \citep{Cantat-Gaudin18}, maximising the following quantity:

\begin{equation}
    \mathcal{L} \propto \prod_{i} P(\varpi_{i}|d,\sigma_{\varpi_{i}}) = \prod_{i} \frac{1}{\sqrt{2 \pi \sigma_{\varpi_{i}}^{2}}} \exp\left( -\frac{(\varpi_{i}-\frac{1}{d})^{2}}{2\sigma_{\varpi_{i}^{2}}} \right)
\end{equation}

where $P(\varpi|d,\sigma_{\varpi_{i}})$ is the probability of measuring a value $\varpi_{i}$ for the parallax of star $i$, if its true distance is $d$ and its measurement uncertainty is $\sigma_{\varpi_{i}}$. This approach  neglects correlations between parallax measurements of all stars, and considers the likelihood for the \epscha~ distance to be the product of the individual likelihoods of its members.

First we checked the distance to the \epscha~ moving group based on known members. Out of 35 stars confirmed by \citet{murphy13}, 32 have distances in the \Gaia\ DR2 catalogue (1$''$ matching radius).  The three missing objects are the members of multiple systems (HD104237 B and C, and $\epsilon$ Cha AB). Additionally, the parallaxes for two sources are much smaller than the average parallax for the sample (mean $\varpi$ = 9.61 mas, with a standard deviation of 1.00 mas; mean errors $\sigma_\varpi$ = 0.10 mas, with a standard deviation of 0.15 mas). For
2MASS J11432669-7804454 (eps Cha 17 from \citealt{feigelson03}), the parallax is $\varpi=5.548\pm0.562$\,mas ($d\approx 181\pm18$\,pc) and the parallax of RX~J1220.4-7407 (eps Cha 36 in \citealt{feigelson03}) is $\varpi=6.714 \pm0.698$\,mas ($d\approx149\pm15$\,pc). Including these sources in the distance estimation does not change the result significantly, nevertheless we suggest re-examining their membership. We obtained a distance of $d=102.2\pm0.4$\,pc when using only members from \cite{murphy13}. When combining with confirmed members from our study, the distance is 102.5 $\pm$ 0.6\,pc. The new distance estimate is somewhat lower than the kinematic distance estimated by \cite{murphy13} ($110\pm7$\,pc), but they stay in agreement within the errors.

%%%%%%%%%%%%%%%%%%%%%%%%%%%%%%%%%%%%%%%%%%%%%%

\section{Conclusions}

In this work we take advantage of \Gaia\ DR2 data to investigate the \cha~complex  and its projected surroundings. From the optical and near-infrared photometry, we selected 35 objects as candidate members for Cha~I (12 candidates), Cha~II (4 candidates) and \epscha\ (18 candidates).
We obtained low-resolution optical spectra for 18 of our candidates, 4 in Cha~I, and 14 in \epscha. Among these we confirmed the young evolutionary stage of 16 candidates.

From \cite{murphy13} we know the membership of 35 objects in \epscha, confirmed based on their kinematics and spectroscopy.
We confirmed 13 new members spectroscopically and rely on X-ray emission as youth indicator for 2 more.
These 15 new members increase the census of \epscha\ by 43\,\%.

Our spectral fits and the location in the HR diagram show that the new members of the \epscha\ moving group have a similar age as the TW Hydrae association (8-10 Myr; e.g. \citealt{ducourant14}).

Unfortunately, the significant increase in statistics does not help to improve our understanding of the origin of \epscha\ and LCC.
\cite{murphy13} already pointed out that with the given similarity in ages, kinematics, and distances, a simple and obvious separation between LCC and \epscha\ may not be possible.
\epscha\ and LCC A0, the southernmost group of LCC, occupy the same proper motion space, and also cannot be distinguished in distance.
The two density  enhancements  seem  to  be  separated by a strip of lower surface density (see Fig.\,\ref{fig:regions}).

However, there is a pattern of several structures from LCC sub-populations (A, B, C, \citealt{goldman18}) and \epscha\ perpendicular to the Galactic plane (based on BANYAN $\Sigma$ model, \citealt{gagne18}, see Fig.\ref{fig:XYZ}). Since the LCC A0 group appears as a bridge between those groups and \epscha\ they may have been born during the same star formation event.

%%%%%%%%%%%%%%%%%%%%%%%%%%%%%%%%%%%%%%%%%%%%%%

\begin{acknowledgements}
    We would like to thank the anonymous referee for their valuable comments that helped improve the manuscript. K.K., K.M., and V.A-A. acknowledge funding by the Science and Technology Foundation of Portugal (FCT), grants No. IF/00194/2015, PTDC/FIS-AST/28731/2017, UIDB/00099/2020, and  SFRH/BD/143433/2019. This publication makes use of VOSA, developed under the Spanish Virtual Observatory project supported by the Spanish MINECO through grant AyA2017-84089.
VOSA has been partially updated by using funding from the European Union's Horizon 2020 Research and Innovation Programme, under Grant Agreement no. 776403 (EXOPLANETS-A).
\end{acknowledgements}
\bibliographystyle{aa}
\bibliography{chamaeleon}

\begin{appendix}

\section{\new{Candidate selection criteria}\label{sec:math}}
{\new Functional forms for the candidates selection criteria presented in the Fig \ref{fig:selection}.}

\begin{table*}
\begin{small}
	\caption{Selection criteria used in this work.}
	\begin{center}
	\begin{tabular}{l l l }
	\hline \hline
			Region & color-magnitude diagram & proper
                        motion ellipse \\
			\hline
Region A & $G_{BP}[mag] = 2.3\cdot(G_{BP}-J)[mag]+ 6.5$&$\frac{(\mu_{\alpha}cos\delta +22.7)^2}{(2.8)^2}+ \frac{(\mu_{\delta}-1)^2}{(3.2)^2}=1$\\
Region B & $G_{BP}[mag] = 2.5\cdot(G_{BP}-J)[mag]+5.7$&$\frac{(\mu_{\alpha}cos\delta +20)^2}{(2)^2}+\frac{(\mu_{\delta}+7.5)^2}{(2)^2}=1$\\
Region C & $G_{BP}[mag] = 2.5\cdot(G_{BP}-J)[mag]+ 4.9$&$\frac{(\mu_{\alpha}cos\delta +41)^2}{(4)^2}+ \frac{(\mu_{\delta}+6.5)^2}{(4.2)^2}=1$\\
			\hline
		\end{tabular}
	\end{center}
	\label{tab:classification_par}
	\end{small}
\end{table*}

\section{Spectral template fits}

\begin{figure*}
   \centering
   \includegraphics[width=\textwidth]{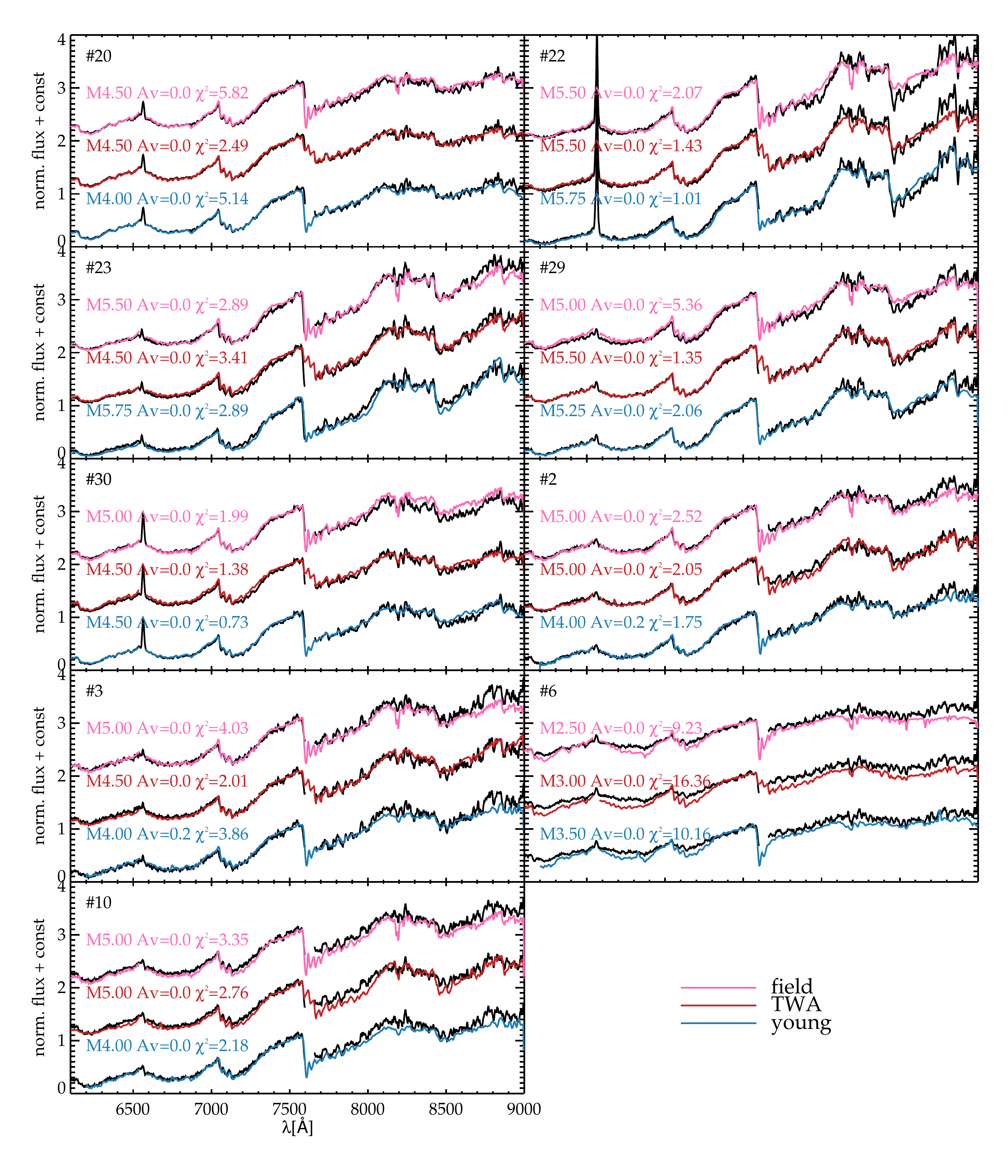}
      \caption{Spectral fitting results for the candidate objects in
        regions A and C (southern portion). The observed spectra are
        shown in black. The best fit young (blue), TWA (red), and
        field (pink) template are shown, along with the corresponding
        spectral type, extinction, and the reduced $\chi^2$ of the
        fit. The H$_\alpha$ emission in some of the template spectra
        has been masked. All the spectra are scaled at 7500\,\AA\ and
        offset for clarity.
      }
         \label{fig:spec_appendix1}
   \end{figure*}

\begin{figure*}
   \centering
   \includegraphics[width=\textwidth]{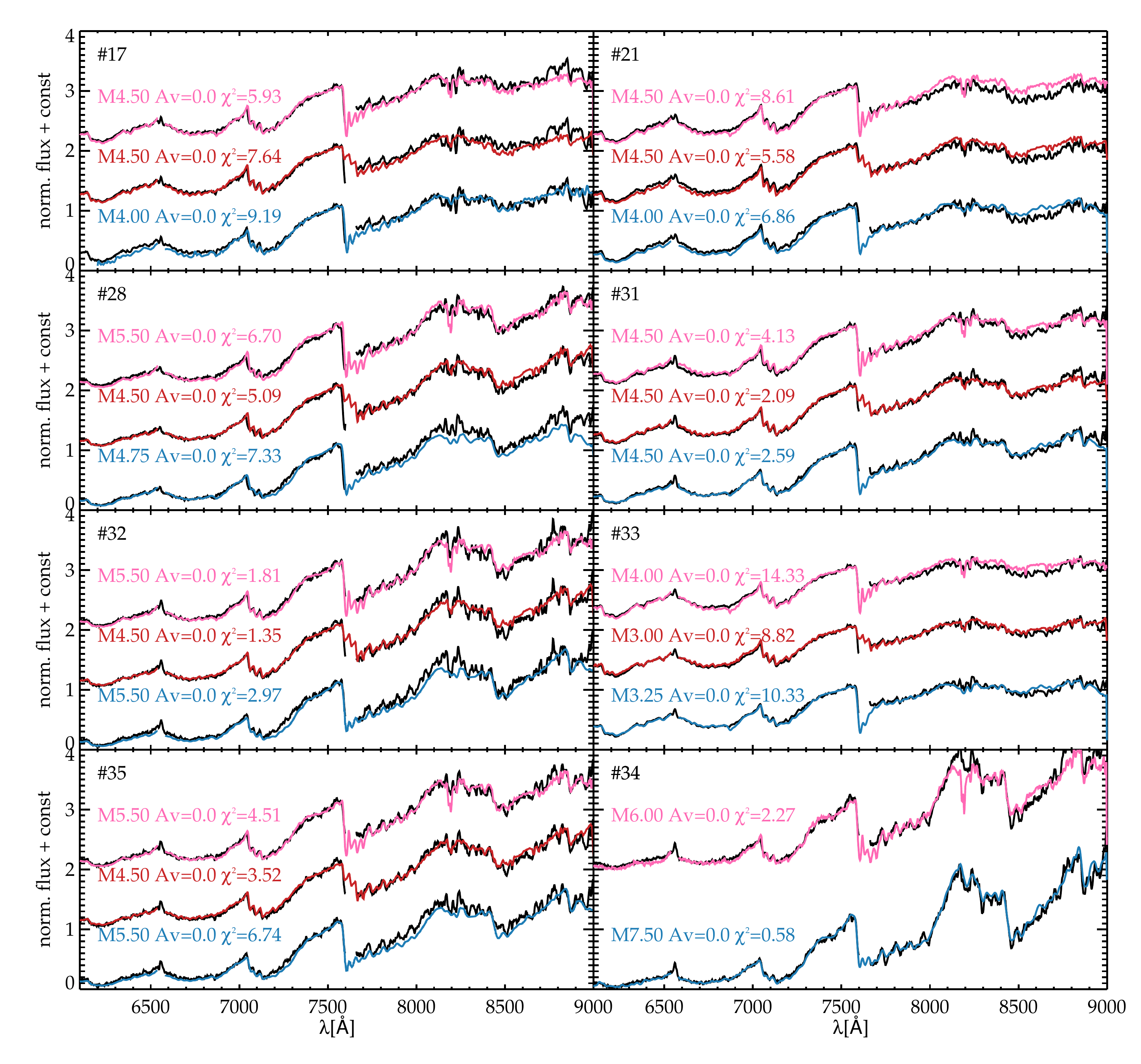}
      \caption{Spectral fitting results for the candidate objects in
        the northern portion of region C. The observed spectra are
        shown in black. The best fit young (blue), TWA (red), and
        field (pink) template are shown, along with the corresponding
        spectral type, extinction, and the reduced $\chi^2$ of the
        fit. The H$_\alpha$ emission in some of the template spectra
        has been masked. All the spectra are scaled at 7500\,\AA\ and
        offset for clarity. For the object $\#34$ we do not show the
        best fit TWA template, since the grid is missing spectral
        types in the range M6-M7.5, inclusive, which coincides with
        the probable spectral type of the object.}
      \label{fig:spec_appendix2}
   \end{figure*}

In Figures~\ref{fig:spec_appendix1} and \ref{fig:spec_appendix2} we show the remaining spectral template fits described in Section~\ref{sec:templates}.

\end{appendix}

\clearpage
\newpage

\end{document}